\documentclass[format=acmsmall, review=false, screen=true, acmjetc]{acmart}
\pdfoutput=1
\usepackage{makecell} 				% to center the heading in table
\usepackage{subfigure}
\usepackage{graphicx}
\usepackage{csquotes}
\usepackage{amssymb}
\usepackage{amsmath}
\usepackage{siunitx}
\usepackage{microtype}
\usepackage{upgreek}
\usepackage{booktabs,tabularx,ragged2e,url}
\usepackage{multirow}
\usepackage{geometry,parskip,setspace,amsmath,amssymb,amsfonts,gensymb}
\usepackage[english]{babel}
\usepackage{fancyhdr}
\usepackage[noend]{algpseudocode}

\usepackage[ruled,linesnumbered,commentsnumbered]{algorithm2e}

\SetAlFnt{\small}
\SetAlCapFnt{\small}
\SetAlCapNameFnt{\small}
\SetAlCapHSkip{0pt}
\IncMargin{-\parindent}

\renewcommand{\arraystretch}{1.3}
\usepackage{xcolor,colortbl}
\definecolor{Gray}{gray}{0.90}
\newcolumntype{a}{>{\columncolor{Gray}}c}
\usepackage{xcolor,soul}
\definecolor{light-gray}{gray}{0.95}
\sethlcolor{light-gray}

\usepackage{enumitem,booktabs}
\usepackage[referable]{threeparttablex}
\renewlist{tablenotes}{enumerate}{1}
\makeatletter
\setlist[tablenotes]{label=\tnote{\alph*},ref=\alph*,itemsep=\z@,topsep=\z@skip,partopsep=\z@skip,parsep=\z@,itemindent=\z@,labelsep=.2em,leftmargin=*,align=left,before={\footnotesize}}
\makeatother

\acmJournal{JETC}
\acmVolume{x}
\acmNumber{x}
\acmArticle{1}
\acmYear{2018}
\acmMonth{5}
\copyrightyear{2018}
\acmArticleSeq{9}

\setcopyright{acmlicensed}
\acmDOI{0000000.0000000}
\graphicspath{{Figures/}}
\newcommand{\eq}[1]{Equation~(\ref{#1})}
\newcommand{\fig}[1]{Figure~\ref{#1}}
\newcommand{\tb}[1]{Table~\ref{#1}}

\setcopyright{acmcopyright}
\acmJournal{JETC}
\acmYear{2018} \acmVolume{1} \acmNumber{1} \acmArticle{1} \acmMonth{1} \acmPrice{15.00}\acmDOI{10.1145/3300971}

\begin{document}
% Title portion. Note the short title for running heads
\title{Neuromemrisitive Architecture of HTM with On-Device Learning and Neurogenesis}

\author{Abdullah M. Zyarah} \author{Dhireesha Kudithipudi}
\orcid{xx}
\affiliation{%
  \institution{Neuromorphic AI Lab, Rochester Institute of Technology}
  \streetaddress{1 Lomb Memorial Dr}
  \city{Rochester}
  \state{NY}
  \postcode{14623}
  \country{USA}}
\email{amz6011@rit.edu}
%\author{Dhireesha Kudithipudi}
% \affiliation{%
%   \institution{Neuromorphic AI Lab, Rochester Institute of Technology}
%   \streetaddress{1 Lomb Memorial Dr}
%   \city{Rochester}
%   \state{NY}
%   \postcode{14623}
\begin{abstract}
\hrule \vspace{2mm}
%Learning transitions of patterns presented to sequence learning algorithms is highly correlated to the success of spatial encoding of sensory information and maintaining its overlap properties.
%The algorithm incorporates both spatial and temporal learning in a unique platform that supports a wide range of applications such as video classification and anomaly detection.
Hierarchical temporal memory (HTM) is a biomimetic sequence memory algorithm that holds promise for invariant representations of spatial and spatiotemporal inputs. This paper presents a comprehensive neuromemristive crossbar architecture for the spatial pooler (SP) and the sparse distributed representation classifier, which are fundamental to the algorithm. There are several unique features in the proposed architecture that tightly link with the HTM algorithm. A memristor that is suitable for emulating the HTM synapses is identified and a new Z-window function is proposed. The architecture exploits the concept of synthetic synapses to enable potential synapses in the HTM. The crossbar for the SP avoids dark spots caused by unutilized crossbar regions and supports rapid on-chip training within 2 clock cycles.  This research also leverages plasticity mechanisms such as neurogenesis and homeostatic intrinsic plasticity to strengthen the robustness and performance of the SP. The proposed design is benchmarked for image recognition tasks using MNIST and Yale faces datasets, and is evaluated using different metrics including entropy, sparseness, and noise robustness. Detailed power analysis at different stages of the SP operations is performed to demonstrate the suitability for mobile platforms.
\end{abstract}

%\{General Terms}: Design, Experimentation
%\{Additional Key Words and Phrases}: CMOS, molecular electronics, nanotechnology
\maketitle
%@@@@@@@@@@@@@@@@@@@@@@@@@@@@@@@@@@@@@@@@@@@@@@@@@@@@@@@@@@@@
% Papers objectives: efficient hardware, optimize the SP, and evaluate the performance, results
\section{Introduction}
%The collective neurons activities, sparse encoding, has recently emerged as a promising information encoding
Mammalian brains process massive amounts of multi-model data for learning, memory, perception, and cognition. All of this information is either spatial, spatio-temporal or spectro-temporal. Modeling such behavior in information processing algorithms can facilitate solutions to complex real-life tasks. Hierarchical temporal memory (HTM)~\citep{hawkins2009sequence,george2009towards} is a theoretical framework that processes spatial and temporal information by emulating the structural and algorithmic properties of the neocortex. HTM offers features such as online learning, multiple simultaneous predictions, sparse distributed representations, and noise robustness~\cite{hawkins2016neurons}. These properties make the algorithm attractive for a wide range of applications such as regression and classification~\cite{HTM_Zeta1_HW_7,melis2009study,xing2012bio}, prediction~\cite{EU_Numbers}, natural language processing and anomaly detection~\cite{anamoly_whitepaper25,lavin2015evaluating}.
At a high level, HTM is a sequence-memory algorithm that learns and recalls patterns of multi-variate time series data. At its core, this is achieved through three key components: encoder, spatial pooler (SP) and temporal memory. The input encoder constitutes the binary distributed representation of input data, whereas the SP and temporal memory continuously transform the input data into sparse distributed representations (SDR) and learn transitions between sequences, respectively.

%Deploying HTM algorithm on mobile devices to bring on-device intelligence is cumbersome. This mainly attributes to the algorithm innate parallel operation and ... which makes it demand massive memory and high computational power~\cite{fan2016hierarchical,melis2009study}.
Deploying the HTM algorithm on mobile and embedded devices can enable real-time prediction and anomaly detection tasks. Specifically, the SP of HTM has critical features including fast adaptation to changing input statistics and noise robustness that can be adopted in hardware. There are few research groups that study the digital and mixed-signal architectures for HTM. However, HTM has been continually evolving and most of the published architectures focus on the earlier deprecated versions of the algorithm. The first wave of architectures were published circa 2007, that focused on the first generation of the algorithm (Zeta). In 2007, Kenneth et al. realized Zeta-HTM on FPGA for image recognition ~\cite{HTM_Zeta1_HW_7}. The model has 81 parallel computational nodes arranged hierarchically in 3 layers and offers 148x speedup over the software counterpart. A Verilog implementation of the single fundamental unit in HTM, a node, is proposed in 2013~\cite{HTM_Zeta1_HW_10}. The second generation of the architectures were investigated circa 2015. Zyarah et al. \cite{zyarah2015reconfigurable}, designed a scalable design with 100 mini-columns and demonstrated for classification with SVM. The authors also proposed a temporal memory design for prediction ~\cite{zyarah2015design}. In 2016, nonvolatile memory based SP implementation is presented by Streat et al. \cite{streat2016non}, considering the physical constraints of the commodity NVRAM. Later, a memristor-based implementation of SP is proposed by James et al. \cite{james2017htm}. Although the proposed design is power efficient, it lacks reconfigurability which is important for learning and making predictions. Recently, Truong et al. presented a memristor-based crossbar to model the SP of HTM algorithm \cite{truong2018spatial}. However, due to the fact that HTM is dominated by dynamic sparse connections, using the traditional crossbar structure leads to \textit{dark spots} (unused regions) in the crossbar. Additionally, most of these research studies do not include the hardware classifier design which is integrated with the SP. Therefore, designing an overarching HTM SP architecture and its associated SDR classifier for energy-constrained platforms with on-device learning supported by dynamic interconnects is still an open research area.

%Having an overarching hybrid design (HTM-SP and its SDR classifier) that gathers the advantages of digital and analog domains in a unified platform while maintaining the required reconfigurability and online learning is still an open research area and has not been demonstrated yet.

%Motivated by the benefits of developing resilient, adaptable, compact, and energy efficient neuromorphic system,

% to emulate the HTM synapses and provides insights about realizing neurons' receptive fields using memristor crossbar.
This paper presents a comprehensive memristor crossbar architecture of the HTM-SP and its associated SDR classifier. The proposed architecture incorporates several unique features that tightly link with the HTM algorithm. A memristor that is suitable for emulating the HTM synapses is identified and a new Z-window function is proposed. The architecture exploits the concept of synthetic synapses to enable potential synapses in the HTM. The crossbar for the SP avoids dark spots caused by unutilized crossbar regions and supports rapid on-chip training within 2 clock cycles.  This research also leverages plasticity mechanisms such as neurogenesis and homeostatic intrinsic plasticity to strengthen the robustness and performance of the SP. The proposed design is benchmarked for image recognition tasks using MNIST and Yale faces datasets, and is evaluated using different metrics including entropy, sparseness, and noise robustness. Detailed power analysis at different stages of the SP operations is performed to demonstrate the suitability for mobile platforms.

The rest of the paper is organized as follows: an overview of HTM is presented in Section \ref{htm_overview}. Section~\ref{design_meth} and~\ref{system_imp} discuss the design methodology and the hardware implementation. The experimental setup and SP evaluation are described in Section~\ref{exper_setup} and ~\ref{sp_metrics}. Section~\ref{exper_results} demonstrates the experimental results. The paper is summarized in Section~\ref{conclusion}.

%@@@@@@@@@@@@@@@@@@@@@@@@@@@@@@@@@@@@@@@@@@@@@@@@@@@@@@@@@@@@
\section{Overview of HTM} \label{htm_overview}
HTM is a sequence memory algorithm that aims at emulating the foundational principles of the neocortex. HTM is structured from ascending hierarchical regions of cellular layers that enable the network to capture spatial and temporal patterns. The cells in HTM are a simplified model of the common excitatory neuron in the neocortex, known as the pyramidal neuron. Similar to pyramidal neurons, HTM cells have hundreds of synaptic connections that enable them to recognize independent patterns of cellular activities. The cell synaptic connections are assigned to three integration zones, namely proximal, basal, and apical~\cite{cui2016comparative,hawkins2004intelligence}. Each zone is composed of either one proximal segment or several dendritic segments. A segment, either proximal or dendritic, comprises multiple synapses to capture the cellular activities of the space to which it is linked. The proximal dendritic segment defines the cell's receptive field in the input space (feed-forward input) and sufficient activities detected on the proximal dendrites lead to the generation of a somatic action potential. The basal and apical dendritic segments hold the synaptic connections with nearby cells and other cells in higher levels in the hierarchy. Therefore, the basal and apical segments are dedicated to observe contextual and feedback inputs. It is important to note that the activities detected on the basal and apical dendrites enable the cells to make prediction via depolarizing it slightly without causing the generation of an action potential~\cite{hawkins2016neurons}.

%The algorithm suggests that the fundamental operations occur in the neocortex involve learning and recalling of temporal sequences.

\begin{figure} [h!tb]
\begin{center}
\includegraphics[width = 0.7 \textwidth]{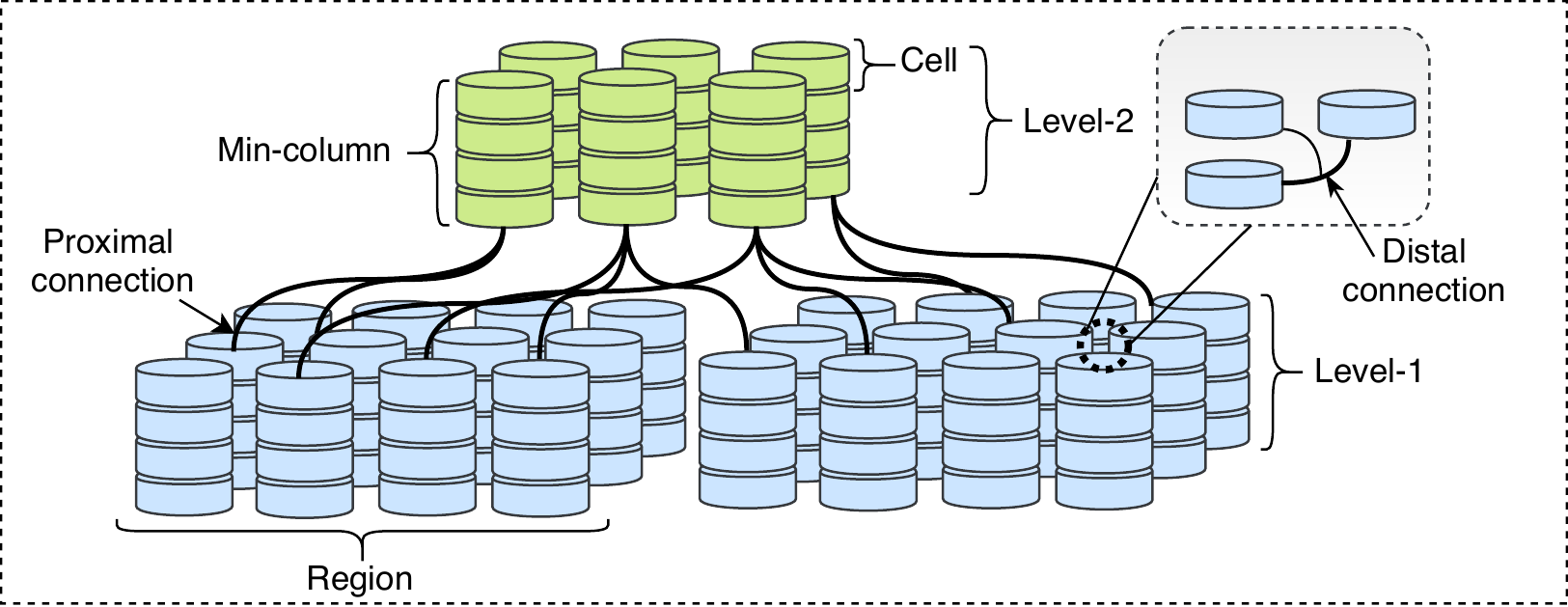}
\caption{High-level architecture of HTM with two levels. The first level has two regions, and one region is confined for the second level. Each region is structured by columns of vertically stacked cells.}
\label{catii}
\end{center}
\end{figure}

% The input data shape the structural connectivity of the network [ref ph]
The cells, in each HTM region, are arranged in a columnar organization called a mini-column. In a given mini-column, cells share the same proximal synaptic connections, i.e. they share the same feed-forward receptive field and stimulated by the same input. Basal segments, on the other hand, allow for the interaction among cells within the same region as such cells learn and recall sequences. The learning in HTM involves adjusting the synaptic connections strength which is defined by a positive scalar value called permanence. However, this process occurs in an online fashion which enables the algorithm to learn not only the spatial features of the input, but also the temporal correlation between them~\cite{padilla2015analysis}. The HTM algorithm is composed of two core phases, namely spatial pooler (SP) and temporal memory, which are discussed in the following subsections:

\subsection{Spatial Pooler Model}
In HTM, learning the spatial patterns in sequential data is performed by the SP. When an input is presented to the network, it gets encoded into a set of sparsely distributed active mini-columns using a combination of competitive Hebbian learning rules and homeostasis~\cite{cui2017htm}. The sparse activation of mini-columns represents the core feature that grants HTM algorithm appealing properties, such as distinguishing the common features between inputs~\cite{SDR_P}, learning sequences, and making simultaneous predictions~\cite{sparse_whitepaper}. Generally, each mini-column is connected to a unique subset of the input space using a set of proximal synaptic connections. When the synapses are active and connected to a reasonable number of active bits in the input space, the proximal dendritic segment becomes active. The activation of the proximal dendritic segment will nominate that mini-column to compete with its neighboring mini-columns to represent the input. By using the $k$-winner-take-all computation principle, the mini-column with the most overlapped active synapses and active inputs inhibits its neighbors and becomes active (winner). The output of the SP is a binary vector, which represents the joint activity of all mini-columns in the HTM region in response to the current input. This binary vector is also known as an SDR vector. The operation of the SP can be divided into three distinct phases: initialization, overlap and inhibition, and learning, as described in Algorithm~\ref{alg:one}.

During the initialization phase (Algorithm~\ref{alg:one}, lines 2-5), which occurs only once, all the parameters of the regions are initialized including mini-columns' connections to the input space, synapse permanences, and boosting factor. Let $S$ be an $n_c\times n_x$ array which holds all the synaptic connections that link $n_c$ SP mini-columns with $n_x-$dimensional input space. Now, let $n_s$ be the maximum number of potential synapses associated with each mini-column and is defined by the non-zero elements in $\vec{s}$ ($\vec{s}$ is a row vector in $S$) whose indexes are generated by a pseudo-random number generator. Similarly, let $\rho$ be a $n_c\times n_x$  array that describes the permanence of the potential synapses in $S$, where the permanence values are randomly initialized with a uniform distribution. After initializing the synaptic connections, the boosting factor for each mini-column is defined to be a scalar value of one. The initialization phase is followed by the overlap and inhibition phase (lines 7-11) in which the feed-forward input is collectively represented by a subset of active mini-columns, namely winning mini-columns. The selection of winning mini-columns occurs after determining the activation level of each mini-column, called overlap score ($\alpha$). The mini-columns' overlap scores for a given region is computed by counting each mini-column's active synapses that associate with active bits in the input space. Mathematically, it is achieved by performing a dot product operation between the feed-forward input vector ($\vec{x}$) and the active synapses vector as in line 9, where the active synapses vector is the result of an element-wise multiplication (denoted as $\odot$) between $S$ and $\rho_*$. $\vec{b}$, here, denotes the boosting factor that regulates mini-column activities.~$\vec{\rho_{*}}$ is a permanence binary vector to indicate the status of each potential synapse, where `1' indicates a connected synapse and `0' an unconnected synapse. Upon the completion of computing the overlap scores, each mini-column overlap score gets evaluated by comparing it to a threshold, known as $minOverlap$ ($O_{th}$) (line 10). The resulting vector ($\vec{e\alpha}$) is an indicator vector representing the nominated mini-columns with high overlap scores. The nominated mini-columns compete against each other with a radius defined by $\xi$ to represent the feed-forward input. Based on the mini-column overlap scores and desired level of sparsity ($\eta$), $n_w$ number of mini-columns will be selected to represent the input, as shown in line 11, where kmax is a function that implements $k$-winner-take-all which returns the top $n_w$ elements within $\xi$. After determining the winning mini-columns ($\vec{\Lambda}$), the learning phase (lines 12-15) starts to update the permanence values of the mini-columns' synapses as necessary, i.e. only the synapses of the active mini-columns are updated. The approach followed in updating the permanence of the synapses is based on the Hebbian rule~\cite{hebb19880}. The rule implies that the connection of synapses to active bits must be strengthened, increase their permanence by $P^+$, while the connection of synapses to inactive bits will be weakened, decrease their permanence by $P^-$, as in line 14, where $\Delta\rho_{j}$ is the change in the permanence array for all mini-columns given an input $\vec{x}$, and $\lambda$ denotes the sum of $P^+$ and $P^-$. After adjusting the synapses permanence, the boosting factor is updated to regulated the activities of the mini-columns, as in line 15, where $\bar{a}(t)$ and $<$$\bar{a}(t)$$>$ contain each mini-column time-averaged activity level and its activity level with respect to its neighbor, and $\gamma$ is a positive constant controlling the adaptation pace~\cite{cui2017htm}.

%describes the initialization of the boosting factor vector which has $n_c$ elements, one scalar value for each mini-column.

% nx: input space size
% ns: number of synapses per columns
% nc: number of columns in a region
% nw: is the number of winning columns to encode each input sample

% Algorithm
\IncMargin{1em}
\begin{algorithm}[t]
\caption{HTM-Spatial Pooling}
\label{alg:one}
\SetAlgoNoLine
\KwIn{$\vec{x} \in \mathbb{R}^{n_x}_{\{0,1\}}$, where $\vec{x} \subset X$ and $X \in \mathbb{R}^{n_x \times n_m}_{\{0,1\}}$ \tcc*[r]{$n_m$: Number of input vectors~~~~~~~}
}
\KwOut{$\vec{w} \in \mathbb{R}^{n_c}_{\{0,1\}}$ \tcc*[r]{$n_c$: Number of columns~~~~~~~~~~~~~}}
\# Initialization: \\
$S_{ind} \sim$ rand.pseudo, where $S_{ind} \in \mathbb{N}^{n_c\times n_s}_{\{1, n_x\}} $ \tcc*[r]{$n_x$: Input vector length~~~~~~~~~~~}
$S[S_{ind}] = 1, \text{where}~S~\text{and}~\rho \in  \mathbb{R}^{n_c \times n_x}$ \tcc*[r]{$n_s$: Number of proximal connections}
$\rho[S_{ind}] \sim$ rand.uniform[0,1] \;
$\vec{b} \in \mathbb{R}^{n_c}$, where $\forall~b[j] = 1$\;
\Repeat{$t > n_m$}
	{
	\# Overlap and Inhibition:\\
 	$\rho_{*}= \text{I}(\rho_ \geq P_{th})$ \;
 	$\vec{\alpha} = \vec{b} \odot \big[(S \odot \rho_{*}) \cdot \vec{x}^T \big]$ \;
 	$\vec{e\alpha} = \text{I}(\vec{\alpha} \geq O_{th})$ \;
 	$\vec{\Lambda} = kmax(\vec{e\alpha}, \eta, \xi)$ \;
 	\# Learning:\\
 	\If{learning = Enable}
 	{
  		{$\Delta\rho = \vec{\Lambda}^T \odot S \odot \rho_{*} \odot \lambda\vec{x} - P^-$\;
  		$\vec{b}= e^{-\gamma(\bar{a}(t) - <a(t)>)}$ \;}

 		}
	}
\end{algorithm}
\DecMargin{1em}

\subsection{Temporal Memory Model}
The main role of the temporal memory is learning sequences and making predictions for future inputs. The cells of the winning mini-columns are involved in this process. The active cells of the winning mini-columns form lateral synaptic connections with the prior active cells, such that cells can anticipate their active state by just examining the distal segments. The number of distal segments that a cell may have depends on the number of distinct patterns that can be predicted. The more distal segments a cell has, the more connections it can have with other cells and thereby more patterns can be predicted. The operation of the temporal memory essentially involves activating the cells of the winning mini-columns to model the input patterns within the context, predicting the future cellular activities, and updating the distal synaptic permanences. As the scope of this paper focuses on hardware implementation of the SP and its SDR classifier, the aforementioned operation will not be discussed. However, a detailed description of the temporal memory and its implementation can be found in~\cite{htm_digital}.

%======================= Design Methodology ========================
\section{Design Methodology} \label{design_meth}
In spite of the fact that the HTM network, in theory, has several hierarchical levels, this aspect has not yet been studied throughly. This work is therefore confined to study and implement only one level/region in HTM. Using one region is equivalent to implementing only the primary sensory region of the neocortex. In the following subsection, modeling of every aspect of the region will be discussed.

\subsection{HTM Synapse Modeling}
HTM cells have a large number of synaptic connections allowing them to detect the pattern of activities occurring in the input space and within the region. Each synaptic level of growth is defined by its permanence value. Typically, the permanence value ranges between 0-1, where 0 indicates the absence of the synaptic connection with a likelihood to form one and 1 indicates the full growth of the synaptic connection~\cite{hawkins2016neurons}. When the permanence value exceeds the threshold, the synapse provides a low-impedance path to the input and vice-versa when the permanence value is below the threshold. However, HTM synapses are binary in nature in the sense that if two synapses permanence exceed the threshold, they exhibit the same properties regardless of their connection strength. While this is the case, the synapse with the highest permanence is harder to forget. In this research, memristor devices are chosen to emulate the synaptic connections in HTM. A memristor is a two-terminal synapse-like nanoscale resistive memory. Its term was coined by Leon Chua in 1971~\cite{chua1971memristor} and the device received rekindled interest when it fabricated by HP labs in 2008~\cite{strukov2008missing}. The device exhibits properties such as low-energy consumption~\cite{prezioso2015training}, small footprint, high integration density~\cite{jo2010nanoscale}, and non-volatility~\cite{borghetti2010memristive}. These features make it an ideal candidate to model the synaptic connections in neuromorphic chips.

The VTEAM memristor model described in~\cite{kvatinsky2015vteam} is used for this research. The device, essentially, is described with two variables: $w$ and $D$ which define the state variable of the device and its thickness. Changing the state of the device, i.e. its resistance value ($R_{mem}$), is considered to have an analog nature. Thus, it is gradual and bounded between the memristor's high resistance state (HRS $\equiv R_{off}$) and low resistance state (LRS $\equiv R_{on}$). The change in the memristor is a function of the voltage applied across the device or the current through it. This work mainly focuses on the voltage driven memristors whose resistance change can be described by~\eq{mem_eq}~\cite{strukov2008missing} and~\eq{delta_w}~\cite{kvatinsky2015vteam}.

\begin{equation}
R_{mem} = \frac{w}{D} \times R_{on} + (1 - \frac{w}{D}) \times R_{off}
\label{mem_eq}
\end{equation}

\begin{equation}
\frac{\Delta w}{\Delta t} =
\begin{cases}
k_{off}.\Big(\frac{v(t)}{v_{off}} - 1\Big)^{\alpha_{off}}.f_{off}(w),&0 < v_{off} < v \\
0, &v_{on} < v< v_{off} \\
k_{on}.\Big(\frac{v(t)}{v_{on}} - 1\Big)^{\alpha_{on}}.f_{on}(w),&v <v_{on} < 0
\end{cases}
\label{delta_w}
\end{equation}

\begin{figure}[h!tb]
\centering
\subfigure[]{\includegraphics[width=45mm, height=35mm]{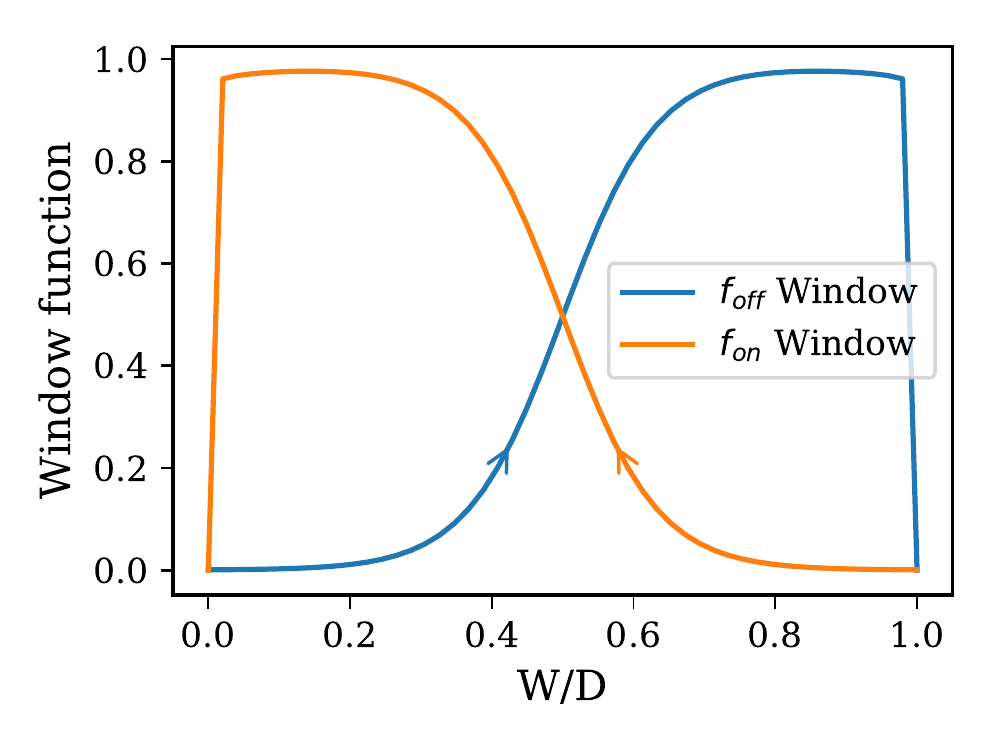}}
%\hspace*{-5em}
\subfigure[]{\includegraphics[width=45mm, height=35mm]{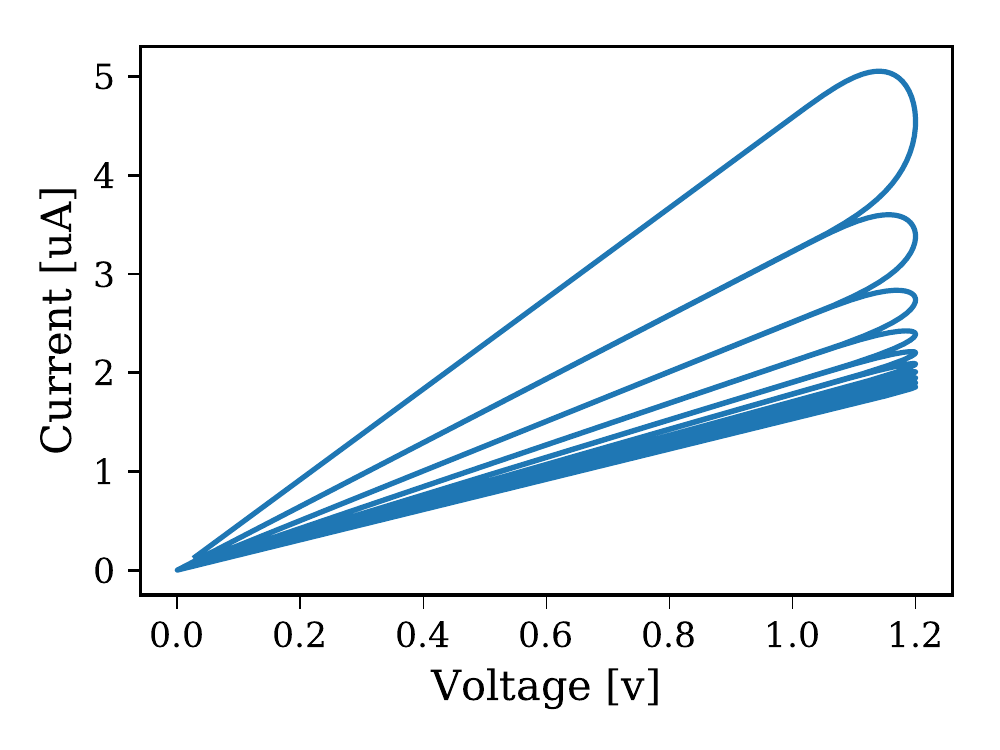}}
\subfigure[]{\includegraphics[width=45mm, height=35mm]{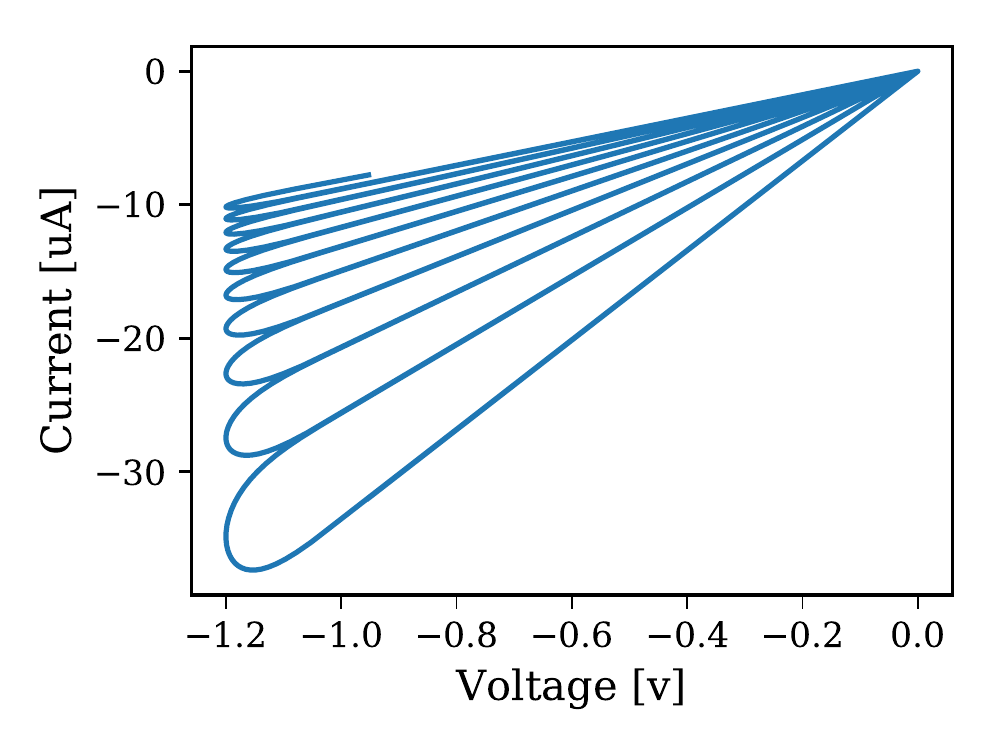}}
\\ \vspace{-1.5mm}
\subfigure[]{\includegraphics[width=45mm, height=35mm]{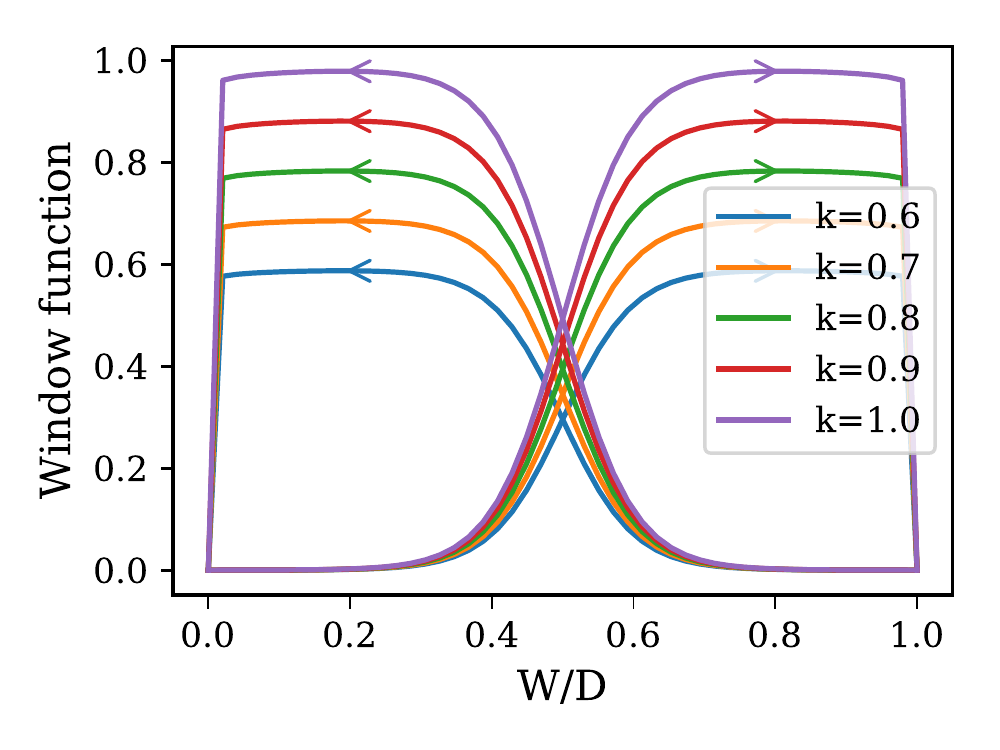}}
\subfigure[]{\includegraphics[width=45mm, height=35mm]{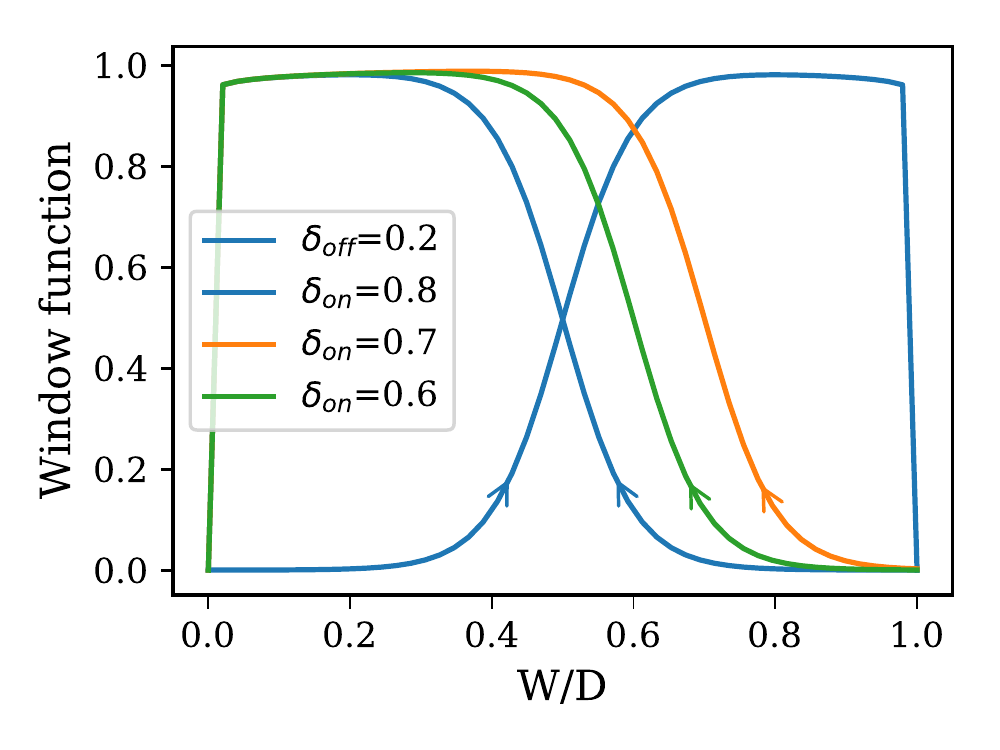}}
\subfigure[]{\includegraphics[width=45mm, height=35mm]{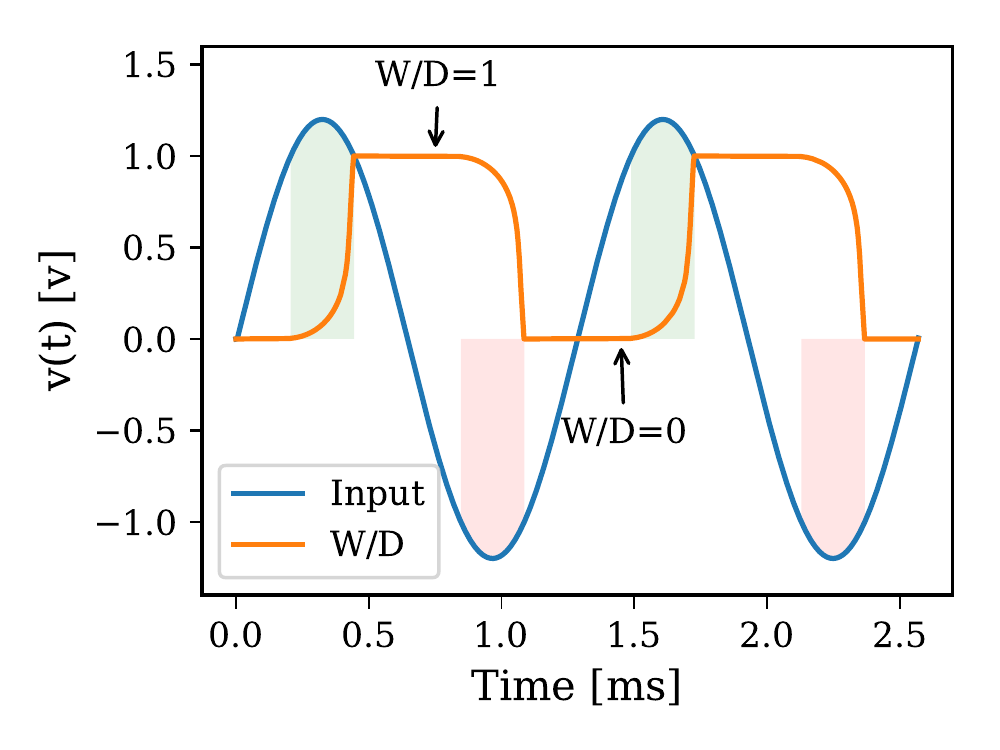}}
\caption{(a) Characteristic curves of the proposed Z-window function to model the HTM synapses behavior. (b) and (c) Hysteresis characteristic curves of the memristor while driving it with a sine wave signal biased with positive and negative DC offset, respectively. (d) A plot shows the linkage to the linear drift model and scalability features, where the arrows to the right refer to $f_{off}$ and the ones to left are for $f_{on}$. (e) Non-symmetrical behavior of $f_{off}$ and $f_{on}$. (f) Modulating the resistance of a memristor device equipped with the Z-window function. The change in memristor resistance is limited to the regions where $|v(t)| > |v_{on}|, v_{off}$, shaded in light green and red.}
\label{window}
\end{figure}

\noindent where $k_{off}$, $k_{on}$, $\alpha_{off}$, and $\alpha_{on}$ are constants, $v_{off}$ and $v_{on}$ are the memristor threshold voltages, and $f_{on}$ and $f_{off}$ describe the device window function. In order to use the memristor device to emulate the HTM synaptic connections, we need to have a memristor that manifests a slight drift when it moves from the boundary toward the mid-point of the device, and as it approaches the mid-point, the drift should be accelerated. In 2016, Brivio et al. proposed a physical memristor that demonstrates properties, which to some extent, match HTM synapse requirements~\cite{brivio2016experimental}. However, modeling this device for circuit simulation requires a special window function so that it exhibits the aforementioned properties. To the best of our knowledge, there is no memristor window function that captures this exponential attribute of HTM synapses. Thus, we developed a window function, called Z-window function, derived from the mathematical formulation of the sigmoid function. The Z-window function has built-in control parameters for adjusting its characteristics and it takes into account the memristor device boundary conditions. Furthermore, it possesses all the attributes of an effective window function such as circumventing the boundary lock problem, providing a linkage with linear dopant drift model, scaling the window function upward and downward~\cite{prodromakis2011versatile, zha2016novel}, and modeling the non-symmetrical behavior of some memristor devices. The proposed window function is given in \eq{win_fun}, where $\tau, \delta$, $k$, and $P$\footnote{Nominal parameters used to achieve most of the plots in~\fig{window} are: $\tau$=15, $\delta$=0.5, $k$=1, and $p$=0.01.} are constants that control the slope of the window function, sliding level (over the x-axis), scalability, and falling slope as it approaches either ends of device terminal, respectively. The subscript $r$ denotes the $on$ and $off$ subscript of the window function. $s(v)$ is a sign function used to make the window function not only depends on the normalized state variable ($\frac{W}{D}$) but also on the voltage across the device and in this case the boundary lock problem is avoided. ~\fig{window} illustrates the window function characteristic curves and its hysteresis\footnote{A sine wave signal has an amplitude of $\pm$1.2v and frequency of 20kHz is used to achieve the hysteresis plots.} as simulated in Cadence using Verilog-A memristor model.

\begin{equation}
f_{r}(w) = \frac{k[s(v) - \frac{w}{D}(-1)^{s(v)+1}  ]^p}{1+e^{\tau(\frac{w}{D} - \delta_{r})(-1)^{s(v)}}}
\label{win_fun}
\end{equation}

\begin{equation}
 s(v)=
\begin{cases}
1,&0 < v_{off} < v \\
0,&v <v_{on} < 0
\end{cases}
\label{sign_win_fun}
\end{equation}

\subsection{Receptive Field}
The receptive field (RF) defines a sub-region in the input space to which a mini-column's proximal synaptic connections are tapped. This section discusses the various approaches of realizing the RFs of the SP mini-columns. It also highlights the advantages of each approach, its constraints and feasibility in realizing a large-scale neuromorphic chip for the HTM algorithm:

\subsubsection{Memristive Crossbar} \label{rig_cross}
\begin{figure} [h!tb]
\begin{center}
\includegraphics[width=80mm, height=42mm]{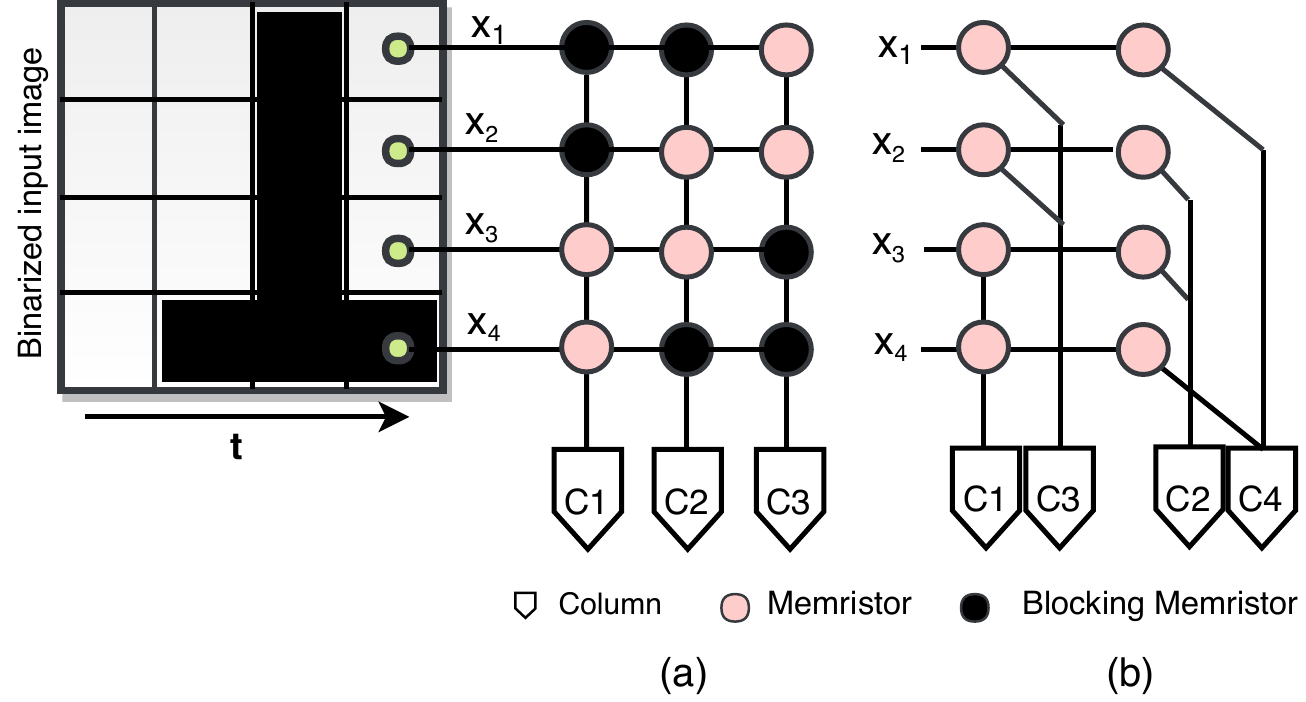}
\caption{Mini-column receptive fields modeled by a sparsely connected memristor crossbar implemented using (a) blocking memristor (b) predefined mini-columns regional connections.}
\label{crossbar_rf}
\end{center}
\end{figure}

The memristive crossbar is mainly composed of perpendicular metal nanowires sandwiching memory elements modeled by memristors~\cite{lu2011two}. The memristive crossbar offers several advantages such as enabling the integration of a large number of memory elements within a compact area and allowing highly-parallel vector-matrix computations. As most neural networks are dominated by vector-matrix multiplications, this makes the memristive crossbar a natural fit for such networks. However, the memristor crossbar structure is really beneficial for densely connected neural networks. When it comes to sparsely connected networks such as HTM, using the crossbar would only be possible by randomly disconnecting devices or setting them to a high impedance state~\footnote{There is another approach proposed in~\cite{cui2016towards} to map a sparse matrix to the crossbar. It is based on decomposing the sparse matrix into a small sub-blocks mapped separately to the crossbar. The sub-blocks with all-zero elements are excluded from the mapping process and therefore reduce the crossbar size. However, such a process requires continuous matrix manipulation and is infeasible for networks with on-chip training. Thus, this approach is not considered here.}. Although both these approaches may result in a sparsely connected crossbar, it is still inefficient modeling. This is because disconnecting devices requires a special burning process, while setting them to a high impedance will not result in perfect current blocking. Having said that, there is a research group that has explored the high impedance method to fulfill a part of HTM's requirements~\cite{truong2018spatial}. The authors suggest using a crossbar in which each column models an HTM mini-column and the rows represent the mini-column's synaptic connections which are connected to the input space. For a given crossbar, the adjacent columns have to maintain a certain level of overlap in the input space.~\fig{crossbar_rf}-(a) shows an example of adjacent columns with two proximal connections each, connected to 4x1 input space (a slice of the presented 4x4 image). Here, it can be noted that the mini-columns $C_1$ and $C_2$ share the input $x_2$ but not $x_3$. In spite of the fact that this method results in partially sparse connections and it enables high-speed computation in the SP, it has several limitations. The first of which is the limited range in the overlap that can be achieved among the neighboring mini-columns because more overlap space implies more unused regions in the crossbar (called the "dark-spot" in the rest of the paper). Second, it leads to current sneak path as the memristors cannot be programmed to zero conductance. Lastly, it lacks the reconfigurability and it makes online learning, which is the most important feature in HTM, more challenging as it requires a training circuit with feedback.

The other possible approach to achieve sparely connected crossbars is based on changing its structure. Instead of using the regular perpendicular cross connections, a regional space to each column is defined such that its connections can be tapped, as shown in~\fig{crossbar_rf}-(b). However, such an approach may have its own challenges during the fabrication process and the same current sneak path issue.

In general, the biggest challenge of adapting the crossbar approach in order to establish the receptive field of each mini-column is the integration between the HTM region and the input space. Using crossbar structure in the ways described above involves establishing hundreds of connections to the input space. This makes HTM architecture over-dominated by the interconnects which eventually lead to undesired noise, scaling limitations, and more power consumption. Furthermore, these connections are rigid in nature and lacks reconfigurability, which is an essential feature to develop an HTM network on chip.
% Note: Regular crossbar implementation lack the overlap receptive field!

\subsubsection{Dynamic Memristive Crossbar}
\begin{figure} [h!tb]
\begin{center}
\includegraphics[width = 0.7 \textwidth]{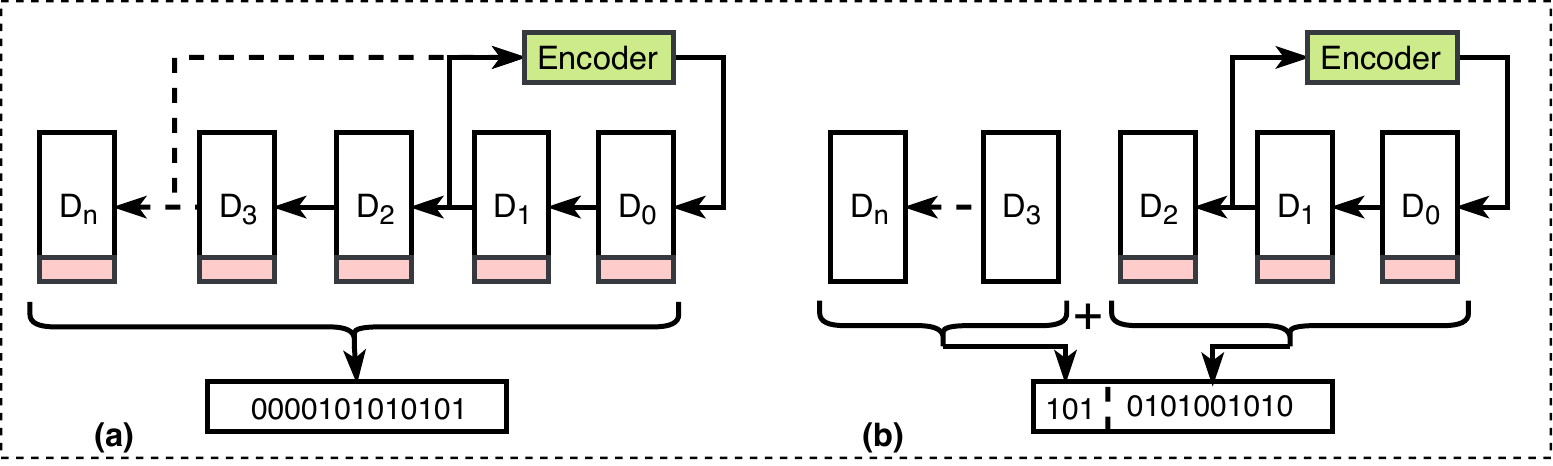}
\caption{(a) LFSR used to generate a global RF. (b) LFSR with partially used registers (red-base) to generate the local RF.}
\label{LFSR}
\end{center}
\end{figure}

The principle concept of this approach is based on using a linear feedback shift register (LFSR) and a memristor crossbar as a single entity to enable crossbar end-terminal reconfigurability\footnote{This is similar to the concept of using synthetic synapses which we proposed in~\cite{zyarah2015reconfigurable}, but here we apply it to crossbar structure.}. Due to the fact that the columns in the crossbar share the rows, a full reconfigurability can only be achieved when the columns are separated to be one-dimensional arrays, where each column models a mini-column in HTM. Each column is assigned its own dedicated LFSR which is initialized by the mini-column index in the HTM region. The RF that is generated by the LFSR can either be local or global. In the global RF, all the registers of the LFSR, shown in \fig{LFSR}-(a), are used to generate random numbers such that the entire input space can be seen by the mini-columns. Given a mini-column, $n_s$ number of potential synapses can be generated by its LFSR to link it with $n_s$ locations in the input space. In the case of the local RF, the LFSR registers are used in a partial manner. Some of them will be used to generate the random numbers whereas the rest are dedicated to provide address shifting. \fig{LFSR}-(b) illustrates the concept of the partially used LFSR. The registers with a colored base represent the registers that will generate the synapses address while the rest are used for shifting. For instance, if an 8-bit LFSR is loaded with a seed of 200, random integer numbers ranged between 192-207 can be achieved if only the 4 least significant bits (LSB) of the LFSR are used.

It turns out that this approach of generating the RF of HTM mini-columns is more expensive in terms of resource utilization and latency in comparison to the rigid memristive crossbar discussed previously in~\ref{rig_cross}. It is, however, more realistic when it comes to scalability because there is no restriction related to the crossbar size, or the number of interconnects being used. Furthermore, on one hand, it satisfies an essential requirement for HTM-SP which involves providing a reconfigurable interconnect that enables implementing topologies of HTM RF, both local and global. On the other hand, it facilitates the communication of the HTM network with the environment and reduces the physical interconnects.

% Another important point that needs to be highlighted is related to the fact that the input to the HTM region is actually sparse in natural and may include hundreds of bits. Establishing connections with all these bits is a real cumbersome  due to the limited packaging IC sizes. Thus, even if crossbar structure is used, special encoder-decoder circuit would be required to bring the feedforward input to the region.

%%%%%%%%%%%%%%%%%%%%%%%%%
\subsection{Homeostasis and Neurogenesis Plasticity Mechanisms}
Homeostasis is an essential mechanism in biologically inspired networks. It prevents neurons from being hyperactive through regulating its threshold of generating a somatic action potential~\cite{zhang2003other}, named $minOverlap$ in HTM theory. The concept of homeostasis in HTM does not involve regulating the $minOverlap$ directly. Rather, it implies exciting the action potential of relatively low active neurons through multiplying the action potential by a positive scalar value called a boosting factor. This results in an effect similar to that of regulating the $minOverlap$ value. The boosting in SP is used to ensure equal likelihood for mini-columns to represent the spatial inputs in SDR form. It is applied through stimulating the mini-columns that have not been active over a predefined time period, i.e not frequently active with respect to its neighboring mini-columns. Consequently, low active mini-columns can have better chance of representing the feed-forward input in future.

\begin{figure}[h!tb]
\centering
\subfigure[]{\includegraphics[width=50mm, height=42mm]{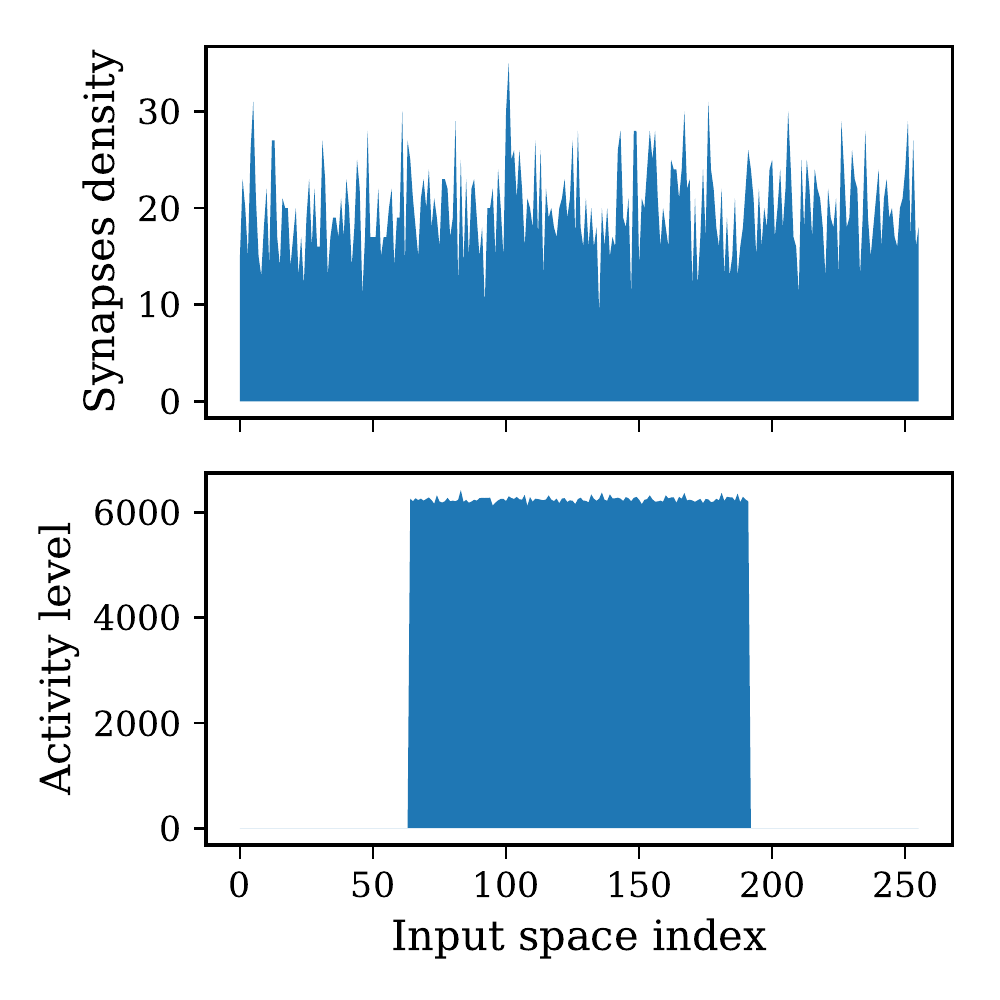}}
\hspace*{1em}
\subfigure[]{\includegraphics[width=50mm, height=42mm]{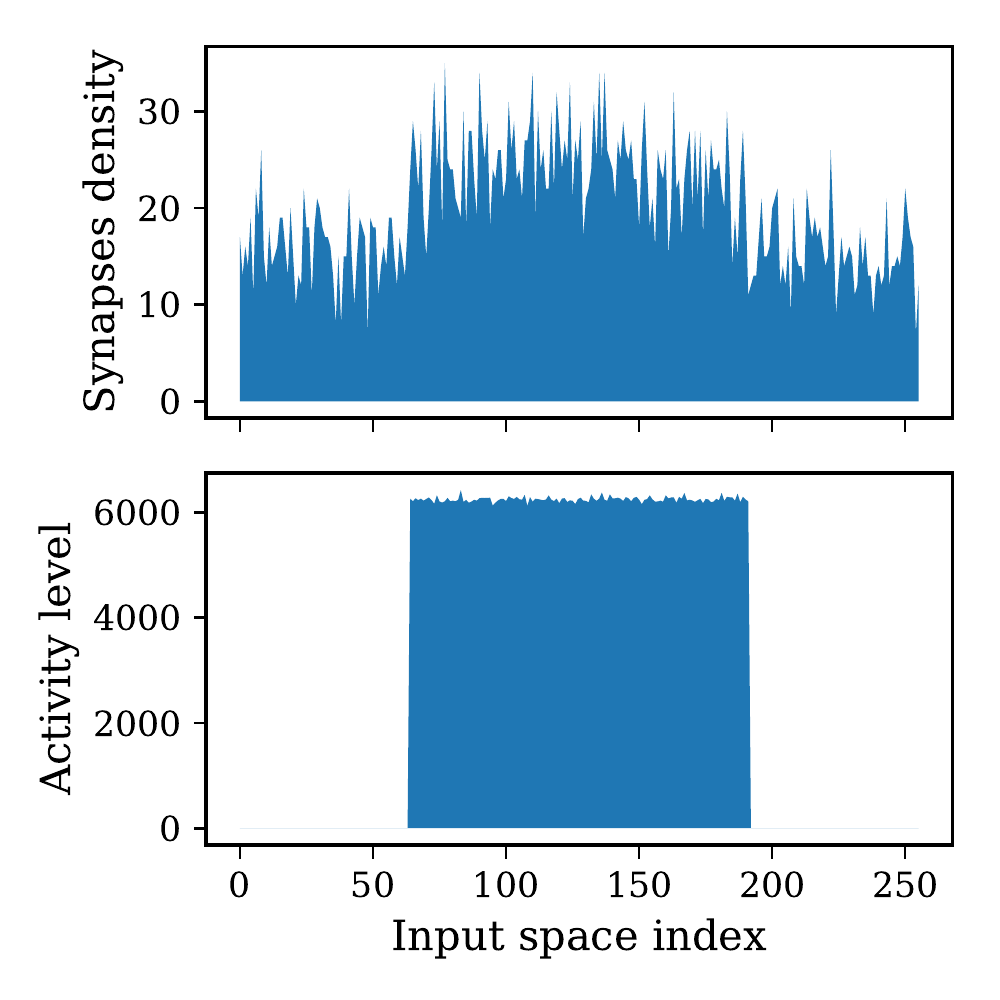}}
\caption{The density of the potential synapses as linked to an input space with activity centered to the middle region when (a) the neurogenesis mechanism is disabled (b) Neurogenesis mechanism is enabled.}
\label{neurogenesis}
\end{figure}

It turns out that using the boosting mechanism is impactful when there is a uniform statistical distribution of information in the input space. Unfortunately, this requirement is not guaranteed especially for visual applications unless a custom encoder is used to process all the inputs. An example of a non-uniform distribution of information in the input space would be the usage of MNIST images. Such non-uniformity in the input space make several mini-columns rarely active or completely inactive. Even the use of boosting here would not cut down the number of inactive mini-columns. Therefore, as a possible solution to overcome this issue, this paper suggests applying the neurogenesis mechanism to HTM. Neurogenesis is a structural plasticity mechanism that suggests 'dead' neurons be replaced with 'new' neurons to enhance network computational capabilities~\cite{souresneural}. Just as in homeostasis, neurogenesis can be applied via tracking the recent mini-column activities over a predefined period of time and comparing it with respect to its neighbor. The mini-columns that were not active frequently are considered 'dead' neurons and should be replaced with new ones. For a given 'dead' neuron, this is achieved by replacing its connections with new randomly initialized connections that are connected to different locations in the input space. Hence, the mini-columns proximal connections will start shifting toward the most active regions in the input space while maintaining a low number of connections to rarely active regions.~\fig{neurogenesis} demonstrates the influence of using neurogenesis on the synaptic connections density in the input space. It can be seen that when the activity in the input space is mediated in the mid-region and neurogenesis is disabled, the connections on the sides are not involved in any computations leading to form non-robust sparse representations. In contrast, when neurogenesis is enabled, the synaptic connections start to move toward the most active spots in the input space and form better representations.

Implementing the neurogenesis mechanism in hardware presents several challenges due to the lack of reconfigurability in interconnects which models the synaptic connections.
%However, such challenges can be overcome if the dynamic memristive crossbar is utilized to model mini-column RFs.
Thus, we adopted the concept of synthetic synapses~\cite{zyarah2015reconfigurable} to enable the reconfigurability in the interconnects and neurogenesis in HTM neuromorphic system. The synthetic synapse concept involves generating the synaptic connections by using an LFSR rather than using rigid connections (just as in dynamical memristive crossbar). By using this, when a given $j^{th}$ mini-column is 'dead' and replaced by $j_{new}^{th}$ 'new' mini-column, all the connections of $j^{th}$ will be removed and replaced by new connections assigned to different locations in the input space and the strength of the new connections are again, randomly initialized.

%The other importing mechanism that we are investigating its effect in this work is the refractory period. Biologically, the refractory period describes the recovery of neuron ionic channels after the generating a somatic action potential. Consequently, it defines the upper limits of neuron firing frequency for a given period of time~\cite{abbott1990model}. Incorporating the refractory period in HTM neuron model is performed by blocking neuron activity for a number of iterations (is chosen to be 5 in this work) and re-enable the neuron again. This results in regular neuron firing and seems to play an a key role in improving network Entropy as will be discussed in the SP metric evaluation section.

%========================= SP Implementation =====================
\section{System Design and Implementation} \label{system_imp}
The high-level architecture of the memristive HTM-SP along with the SDR classifier is shown in~\fig{sp_fig}-(a). The SP architecture comprises a set of mini-columns to spatially process the feed-forward input and a main control unit (MCU) to enable the mini-columns to interact with the ambient environment. When the input is presented to the network, the MCU relays it to the mini-columns such that each mini-column's active proximal connections will be identified and counted. The mini-columns with active proximal segments compete against each other and the top X\% mini-columns that receive most of the input are activated. The selection of the top X\% mini-columns is performed using a voltage-mode winner-take-all (WTA) circuit. After identifying the best mini-columns that represent the input, the learning process starts. The learning process occurs in an online fashion which grants the network the ability to adapt continuously to the input changes. Learning is governed by a Hebbian rule and involves modulating the connections for the active columns only. Upon the completion of the learning, the output of the SP is generated and then passed to an SDR classifier when the network is used for classification applications (see ~\fig{sp_fig}-(b)). In the following subsections more details about the implementation of the SP and the SDR classifier will be provided:

\begin{figure} [h!tb]
\begin{center}
\includegraphics[width = 110mm, height = 60mm]{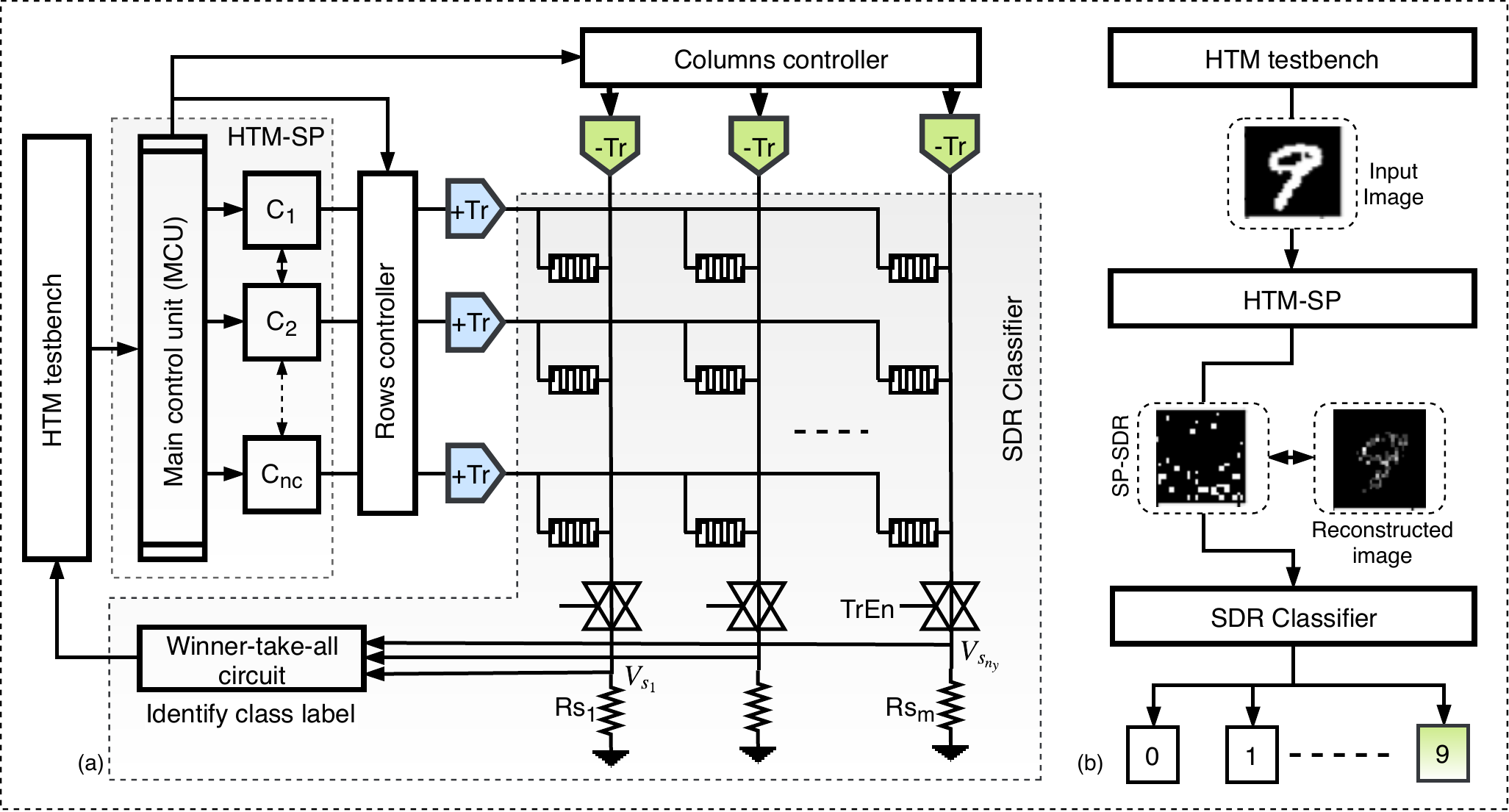}
\caption{(a) The high-level architecture of the proposed design, HTM-SP for sparse representation of the feed-forward input, the SDR classifier to recognize the SDR representations, and training circuity to enable the learning for both SP and SDR classifier. (b) High-level diagram demonstrating the input images after being processed by each stage in the network.}
\label{sp_fig}
\end{center}
\end{figure}

\subsection{HTM Spatial Pooler}
The SP is essentially composed of an MCU and mini-columns. As aforementioned, the MCU bridges the mini-columns with either the sensory input or other regions in the hierarchy and the mini-columns are the units where the main SP computations occur. In the next subsection, the mini-column circuit, its training circuity, and the WTA unit are discussed in more details.

\subsubsection{Mini-Column}
\begin{figure} [h!tb]
\begin{center}
\includegraphics[width = 100mm, height = 40mm]{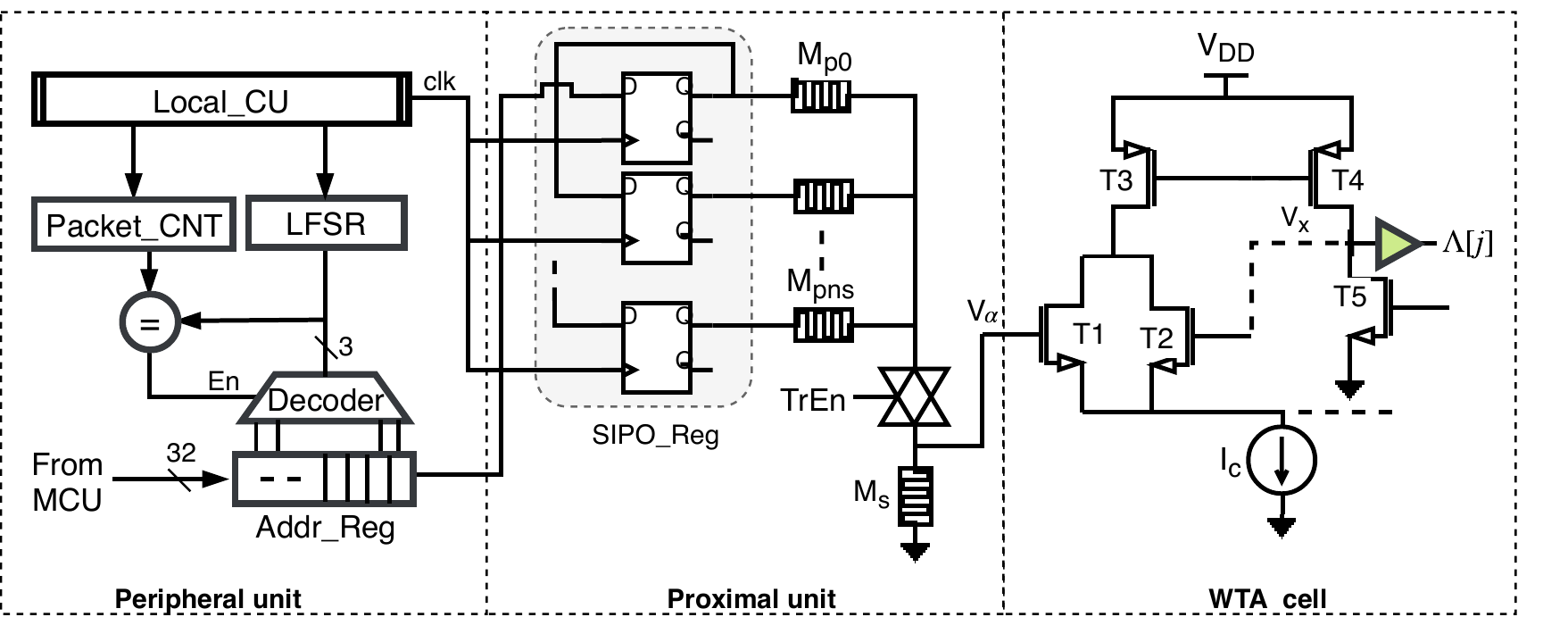}
\caption{The circuit diagram of a mini-column in an HTM region. It consists of a peripheral unit in which the proximal connections are generated and connected to input space, a proximal unit to store the connections strength, and a WTA cell to enable the mini-columns to compete for input representation.}
\label{column}
\end{center}
\end{figure}

\fig{column} depicts the architecture of the SP mini-column. The mini-column is modeled by three units named: peripheral unit, proximal unit, and WTA cell. The peripheral unit models the part of the mini-column in which the proximal connections are generated and connected to the input space. The proximal unit and WTA cell hold the proximal connection permanences and a contesting unit that enables each mini-column to compete with its neighbors for the input representation, respectively. The input to the mini-column is generated by the network encoder (modeled by the network testbench). The encoder task, in this work, is confined to binarizing the images and slicing them into small patches to minimize data movement and the required storage units. Sequentially, each patch is presented to the mini-column and stored into an $Addr\_Reg$. Meanwhile, the LFSR generates a random number indexing the observe patterns activities in the feed-forward inputs.

Given an input image of size 32x32, it is sequentially fetched to the network in the form of patches, where each patch is a 32-bit row vector. When the input patch is stored in the $Add\_Reg$ and the LFSR generates an address for a location in the received patch, a matching score is stored in a synapses register which is modeled by $n_s \times 1$ serial-in-parallel-out shift register. Once all inputs are received, the output of the synapses register is fetched to the word-line of $n_s \times 1$ memristive crossbar where the proximal synapse permanences are stored. The input voltages to the crossbar will be converted in form of current through the memristor and the output is collected at the crossbar bit-line. The output of the crossbar which modulates the mini-column overlap score to current is then boosted. Boosting is done via the usage of a sense memristor ($M_s$). The boosting factor is inversely proportional to $M_s$ ($\equiv \frac{1}{g_s}$) conductance as such decreasing $g_s$ value leads to increase the boosting factor and vice versa. The output at this point, $V_{\alpha_j}$, is given by~\eq{overlap_score}:
\begin{equation}
V_{\alpha_j} = \frac{\sum\limits^{n_s}_{i=1} g_i~V_i }{g_s + \sum\limits^{n_s}_{i=1} g_i}
\label{overlap_score}
\end{equation}
\begin{equation}
V_{\alpha_j} \approx \frac{1}{g_s}\sum\limits^{n_s}_{i=1} g_i~V_i,~~~~\text{if}~~ \frac{gs}{\sum\limits^{n_s}_{i=1} g_i} >> 1
\label{overlap_score2}
\end{equation}

where $g_i$ ($\equiv \rho_{i,j}$) indicates the conductance of $i^{th}$ memristor and $V_i$ is the $i^{th}$ input voltage. $V_{\alpha_j}$ ($\equiv \alpha_j$) denotes the $j^{th}$ mini-column overlap score. Upon the completion of computing the overlap score, its value, which is sampled by the sense memristor, is then presented to a WTA circuit. The WTA performs a $kmax$\footnote{The function $f (\equiv kmax)$ is computed in the next winner-take-all, section (\ref{wta_sec}).} operation on $V_{x_j}$ followed by a thresholding, to generate the final $j^{th}$ mini-column output, ($\Lambda_j$), as given in \eq{col_out}.
\begin{equation}
\Lambda_j =
\begin{cases}
1 , & ~V_{x_j} > V_{th},~where~V_{x_j} = f(V_{\alpha_j}) \\
0,  & Otherwise
\end{cases}
\label{col_out}
\end{equation}

The minimum input to the WTA circuit should be no less than 0.2v so that a cell is activated. This requirement implicitly realizes the concept of $minOverlap$ in HTM which implies that mini-columns overlap score should be large enough to enable it for competing against other mini-columns to represent the input. The output of the WTA cell indicates the mini-columns status, where logic '1' refers to a winner. Selecting the winners is followed by the learning process in which each winning mini-column synapses are adjusted in response to the stimulated feed-forward input and according to Hebbian learning rules.

%According to the Hebbian rule, the synapses that were stimulated by active bit are enforced by increasing their strength while the ones connected to inactive bit are weakened. In this manner, the mini-columns will have more likelihood to be active for the inputs that have seen before.

%In the following subsections, the WTA circuit and learning in mini-columns will be discussed in more details:

%Notes: the Rs resistance has to be way more than that of training transmission gate so that most of the current flows toward the transmission gate. However, not quite sure how would the current effect the training process if it passes toward the RS resistance. This requires further analysis.

%%=====================
\subsubsection{Winner-take-all Circuit} \label{wta_sec}
\begin{figure} [h!tb]
\begin{center}
\includegraphics[width = 110mm, height = 45mm]{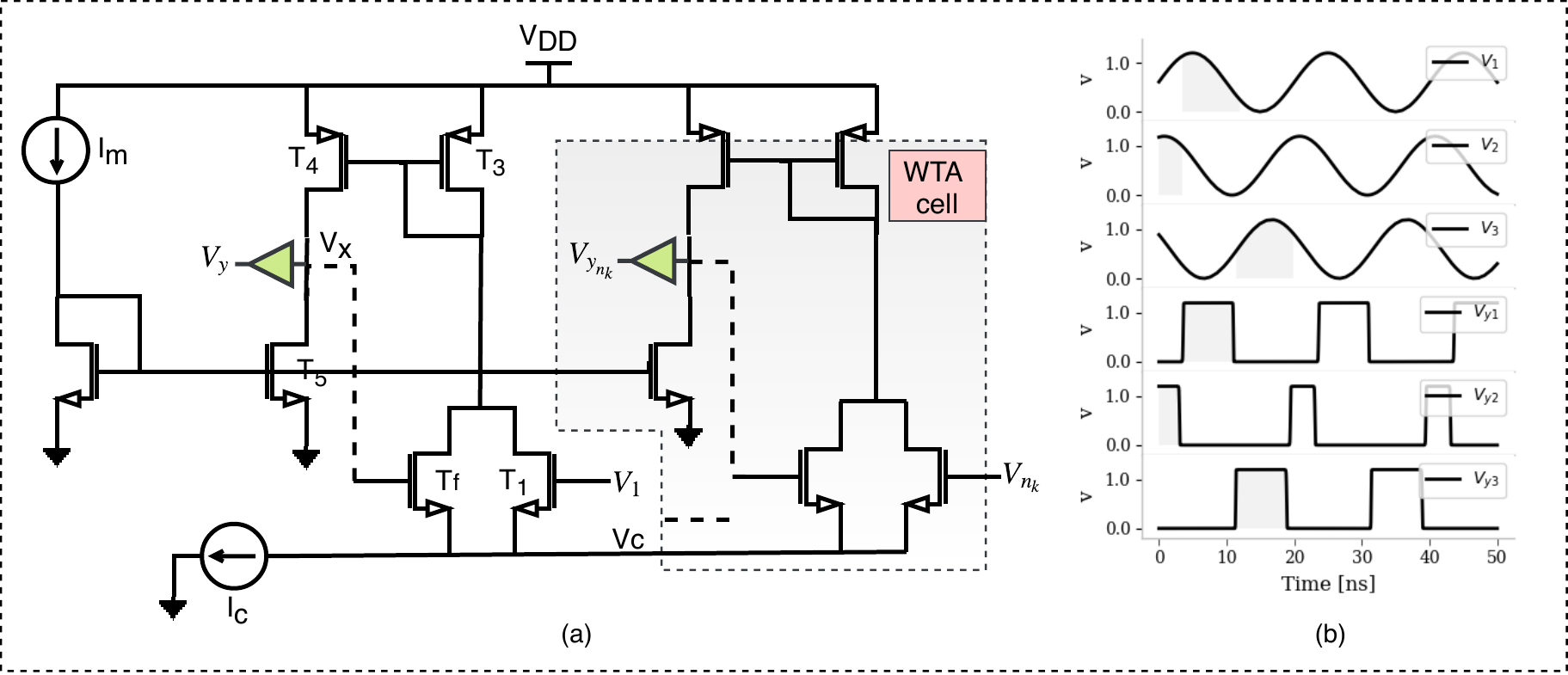}
\caption{(a) WTA circuit ($n_k$ number of cells) with local excitatory feedback., (b) A waveform diagram demonstrating the WTA circuit operation while presenting three sine waves with 75$\degree$ phase shift between them as inputs. The input with maximum amplitude for a given period is shaded with light gray.}
\label{WTA}
\end{center}
\end{figure}

The WTA cells are utilized as apart of the mini-columns circuit to select the winners in each local (or global) cluster and in the SDR classifier to identify the winning class labels. \fig{WTA}-(a) depicts the WTA circuit which models a simple local competitive algorithm which is naturally imposed through Kirchhoff current law (KCL). Each branch in the circuit has an NMOS transistor ($T_1$) to capture the input signal ($= V_\alpha$ in the mini-column circuit and $V_s$ in the SDR classifier) of one competitor. The competitors interact with each other through the shared point $V_c$. When inputs are presented to the circuit, the potential of $V_c$ follows the input with the highest voltage and turns off all the other transistors. The cell conveying most of the bias current, $I_c$, is identified as a winner. Given that all the transistors operate in subthreshold regime, applying an input voltage $V_{G_j}$ at the gate of the transistor in the $j^{th}$ branch results in a current $I_j$, which can be approximated by \eq{subth_cur}~\cite{razavi2017design}:
\begin{equation}
I_{j} = I_o~(\frac{W}{L})~e^{\frac{V_{GS_j}}{nU_T}}
\label{subth_cur}
\end{equation}

\noindent where $I_o$ is the zero-bias current for the given device, $\frac{W}{L}$ is the transistor channel width to length ratio,  $U_T$ and $n$ indicate the thermal voltage and the subthreshold slope coefficient, respectively. For the given circuit with $n_k$ branches, according to KCL, the branches' current should sum up to $I_c$, as given by \eq{sum_current}. By using \eq{subth_cur} and \eq{sum_current}, we can solve for the current flowing in each branch as shown in \eq{branch_current}, which is identical to the softmax function.
\begin{equation}
I_c = \sum^{n_k}_{j=1} I_j
\label{sum_current}
\end{equation}
\begin{equation}
I_j = {I_c}~\frac{e^{\frac{V_{G_j}}{nU_T}}}{\sum\limits^{n_k}_{j=1} e^{\frac{V_{G_j}}{nU_T}}}
\label{branch_current}
\end{equation}

Recall that the output of the SP is a voltage and is represented in a binary sparse form. Thus, we designed the WTA to be a voltage mode circuit. In order to maintain the same normalized exponential relationship between the input and output (described in \eq{branch_current}), the current in each branch is sent to a current comparator via a current mirror formed by $T_3$ and $T_4$, as shown in \fig{WTA}. The mirrored current is compared to a fixed reference current resulting in a voltage drop across the point $V_{x_j}$, which can be calculated using~\eq{vx_eq}:
\begin{equation}
V_{x_j} = \frac{1}{\lambda_5} \Big[\frac{2Ai~I_{3}}{\beta_5 (V_{GS_5} - V_{th_5})^2} - 1 \Big]
\label{vx_eq}
\end{equation}

\noindent where $\lambda$ is the channel-length modulation, $\beta$ is the transconductance parameter, $Ai$ and $V_{th}$ denote the current mirror gain between $T_3$ and $T_4$ and transistor threshold voltage. By substituting \eq{branch_current} in \eq{vx_eq}, the output node, $V_{x_j}$, is calculated in \eq{vout_eq}:

\begin{equation}
V_{x_j} =  \psi_5{I_c}~\frac{e^{\frac{V_{G5_j}}{nU_T}}}{\sum\limits^{n_k}_{j=1} e^{\frac{V_{G_j}}{nU_T}}} - \frac{1}{\lambda_5}
\label{vout_eq}
\end{equation}

Due to the fact that $\psi_5$ is approximately constant and is given by $\frac{2Ai}{\lambda_5\beta_5 (V_{GS5} - V_{th5})^2}$, \eq{vout_eq} indicates that the output voltage $V_{x_j}$ for branch $j$ has a normalized exponential relationship with the input $V_{G_j}$. Such relation has a unique benefit for WTA circuit because it maximizes the difference between the inputs. It generously rewards the input with the highest value and punishes the losing ones.
%Furthermore, this mechanism makes the power consumption of the circuit extremely low.
Most of the power consumption is dominated by the winning cells which are low in number compared to losing cells.
% The reason we design a WTA circuit:
% \begin{itemize}
% \item To generate the output in form of voltage
% \item Developing a scalable WTA circuit which can enable k-number for winner and this eventually provides for more flexibility when it comes to sparsity level tuning
% \item The circuit offers low latency, high resolution, and low power consumption.
% \end{itemize}

It is important to notice here that unlike most other WTA circuits, in literature~\cite{lazzaro1989winner, ramakrishnan2014vector, kulej2017sub}, all outputs are buffered to provide enough driving capabilities when transmitting signals across long distances. Also, few of the previous WTA circuits are endowed with a hysteresis mechanism to increase network stability and prevents the selection of a potential winner unless they are strong. Due to the fact that the hysteresis is achieved via a local excitatory feedback, some of these circuits require a reset process to any competition as in~\cite{fish2005high}. In the proposed WTA circuit, the hysteresis characteristic is introduced via the positive feedback formed by the transistor $T_f$. Additionally, having a current comparator improves the stability further as it imposes a threshold current that needs to be crossed to switch cells status. The other advantage of using the current comparator is that it enables more than one winner, which is a desirable feature especially in HTM as it allows controlling the network output sparsity level.

%%%%%%
\subsubsection{Mini-column Learning Circuit}
The learning in HTM is performed in an online fashion and it involves modulating the synaptic permanence of the winning mini-columns only. As aforementioned, the proximal synaptic connections of each mini-column are emulated by one-dimensional memristive array. Therefore, the training here can be performed simultaneously. By using a modified Ziksa unit~\cite{zyarah2017ziksa}, training each mini-column synaptic connections can be performed in two clock cycles. After computing the mini-column overlap scores, the synaptic connections that were connected to active bits in the input space has their D-Flip-Flop (DFF) set to high and vice-versa for the synapses connected to inactive bits. All the DFF outputs are buffered with a modified NOT gate that generates a logical level output during the normal operation and a training voltage during the learning phase. When the $TrEn$ signal is generated (active-low), the positive terminals of the memristors will be connected to $V_{Tr}$ if the output of the DFF is high and $GND$ otherwise. The other terminal of the memristor will be controlled by $Tr_1$ and $Tr_2$. During the first cycle of training, $Tr_1$ is set to ON by $Tune^+$. If DFF output is low, this causes a voltage drop across the memristors that needs to be adjusted to exceed the threshold leading to an increase in its resistance. During the second clock cycle, the same procedure will be applied but in the opposite manner\footnote{The training circuit is verified while considering memristor device variability for resistance range and threshold voltage. Variability with normal distribution has been considered with mean defined by device parameters and standard deviation (STD) equals to 10\% of the mean for resistance and 5\% of the mean for threshold voltage.}.

\begin{figure} [h!tb]
\begin{center}
\includegraphics[width = 110mm, height = 48mm]{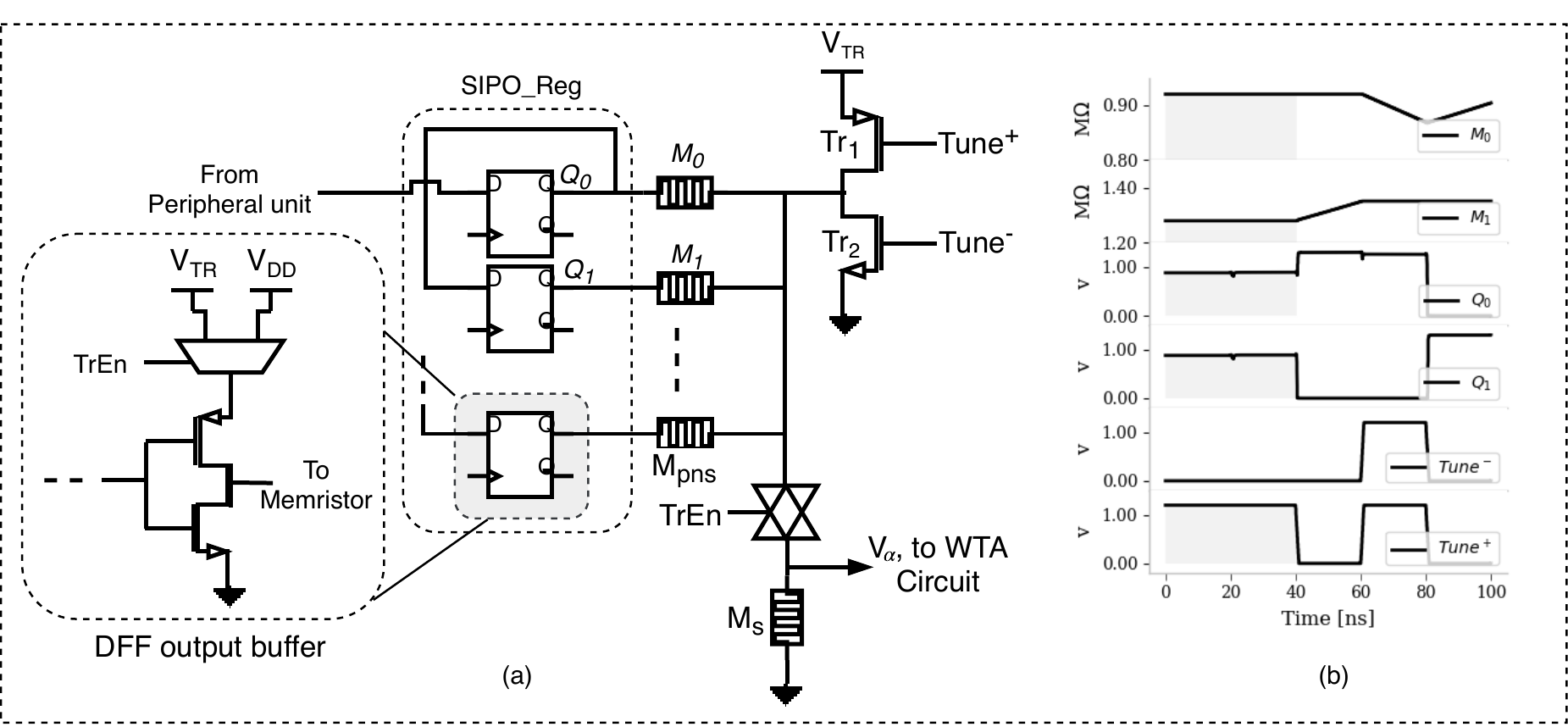}
\caption{(a) The training circuit of the proximal synaptic connections in an HTM mini-column, (b) A waveform diagram demonstrating the operation of the training circuit during the testing period (shaded in light gray) and training period.}
\label{column_tr}
\end{center}
\end{figure}

\begin{figure} [h!tb]
\begin{center}
\includegraphics[width = 100mm, height = 40mm]{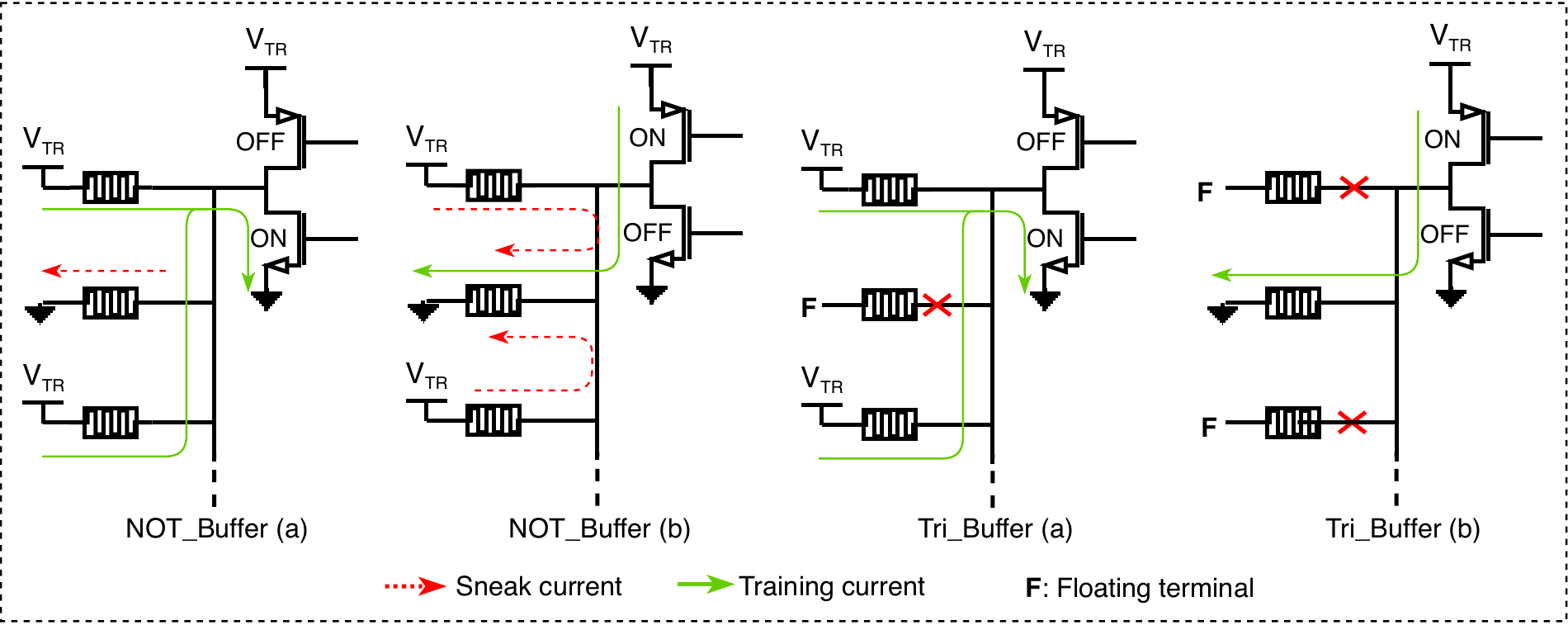}
\caption{The possible scenarios for the current sneak paths when a DFF is buffered with (a) A NOT gate, (b) A Tri-state buffer to drive the proximal connection memristors.}
\label{sneak}
\end{center}
\end{figure}

The downside of using the inverters of DFFs in conjunction with Ziksa unit is that the network will suffer from the sneak path issue especially during the learning phase. However, this issue can be overcome by buffering the output of DFF with a tri-state buffer rather than an inverter gate. Using a tri-state buffer allows the memristors that are not involved in the training process~\footnote{The memristors that are not involved in the training are those that need to be decremented during the first clock cycles or those that need to be incremented in the second clock cycles of training.} to be floating such that it does not draw any current.~\fig{sneak} covers the possible scenarios for the sneak paths when a NOT and tri-state buffers are used.

\subsection{SDR classifier}
The SDR classifier recognizes the SDR combinations as generated by the SP and generates the predicted class label. It is built using a softmax circuit ($\equiv$ WTA circuit shown in~\fig{WTA}) with $n_y$ number of units, where $n_y$ denotes the number of class labels that needs to be recognized. All the SDR classifier units are interconnected with the SP mini-columns in a dense manner through weighted connections, $\varpi \in \mathbb{R}^{n_c \times n_y}$. Initially, the weighted connections of the classifier are randomly initialized and then tuned according to the delta rule, given in \eq{delta_rule}, where $\varpi_{j,k}$ is the weight of the connection between the $k^{th}$ classifier unit and $j^{th}$ SP mini-column, $\sigma$ is the learning rate. $\vec{\Lambda_j}$ denotes the $j^{th}$ mini-column output, and $y^*_k$ and $y_k$ are the predicated and expected class labels, respectively. It is important to mention here that the delta rule can be applied using Ziksa training circuitry as described in~\cite{zyarah2017ziksa}.

\begin{equation}
\label{delta_rule}
    \Delta \varpi_{k,j} = \sigma \times \vec{\Lambda_j} \times (y^{*}_{k} - y_{k})
\end{equation}

\begin{equation}
y^*_k = \frac{e^{\vec{\Lambda} \times \vec{\varpi_i}}}{\sum\limits^{n_y}_{i=1} e^{\vec{\Lambda} \times \vec{\varpi_i}}}
\label{sdr_Vo}
\end{equation}

%The SDR classifier utilized in this work is memristor based and trained with the maximum likelihood learning rule. More details about the columns architecture and the learning process is provided in the following subsections:

%========================= Network Parameters ===============================
\section{Experimental Setup} \label{exper_setup}
\subsection{Device Parameters}
The Verilog-A memristor model, VTEAM~\cite{kvatinsky2015vteam} is mapped to the physical memristor device by Brivio~\cite{brivio2016experimental}. The device resistance range is opted to fulfill the design constraints and to ensure proper operation. The low resistance state (LRS $\equiv R_{on}$) is chosen to be 200k$\Omega$ and the high resistance state (HRS $\equiv R_{off}$) is 5M$\Omega$ such that, sufficient amount of input current causes a voltage drop of $\approx$0.85v\footnote{This is when a mini-column overlap score exceeds $minOverlap$} across the sense memristor. This range of memristor resistance also  minimizes the power consumption of the crossbar array. Given the technology node and the supply rail, from the mini-column circuit (\fig{column}), the sense memristor range can be estimated as in \eq{ms_min}\footnote{Parameter to find $M_{s_{min}}$: $g_i$=2e-7$\frac{1}{\Omega}$, $n_s$=32, $V_s$=0.2v, $\forall V_{m_i} = 0.9v$. Parameters to find $M_{s_{max}}$: $g_i$=5e-6$\frac{1}{\Omega}$, $n_s$=24, $V_s$=0.9v, $\forall V_{m_i} = 0.9v$}. Here, the minimum value of the sense memristor is chosen such that it does not bring $V_\alpha$ to more than 0.3v even when the inputs are set to high. On the contrary, activating three-quarter of the proximal connections is enough to bring the voltage drop across the sense memristor to maximum ($\approx$0.85v).

\begin{equation}
M_s = \frac{\sum^{n_s}_{i=1} M_i}{ \Big[\frac{\sum\limits^{n_s}_{i=1} V_{m_i}}{V_\alpha} -1 \Big]}
\label{ms_min}
\end{equation}

\begin{table}[h!tb]
\small
\caption{The device parameters used to simulate the SP and its SDR classifier.}
\label{hard_para_table}
\setlength\tabcolsep{3 pt}
\renewcommand{\arraystretch}{1.3}
\begin{center}
\begin{tabular}{|l|c|}
  \hline
\rowcolor{Gray}
\textbf{Parameter} & \textbf{Value} \\ \hline
  Proximal memristor range    & 200k$\Omega$ - 5M$\Omega$ \\ \hline
  Memristor threshold               & $\pm$1v \\ \hline
  High-Frequency clock       		& 50 MHz \\ \hline
  Training voltage                  & $\approx$1.2 v \\ \hline
  Sense memristor range             & 20k$\Omega$-80k$\Omega$ \\ \hline
\end{tabular}
\end{center}
\end{table}

Recall that the proposed HTM-SP network offers online learning. However, in order to test the system for generalization, the learning might be turned off. Since a threshold voltage memristor model is adapted, a threshold voltage of $\pm$1v is used to disable the device undesired adjustment when the learning is disabled. Upon the completion of setting the mini-columns' parameters, the system parameters are defined such that system offers real-time data processing, online learning, and minimum power consumption. The system parameters are reported in~\tb{hard_para_table}.

%The system parameters here are represented by the system voltage rail which is constrained by the used technology node and is chosen to be 1.2v. The other parameter is the system clock frequency which is set to 50 MHz.  All the parameters used during the simulation are reported in~\tb{hard_para_table}

\subsection{SP Network Setup}
In order to set the optimal network parameters for the utilized verification benchmarks, the particle swarm optimization (PSO)~\cite{eberhart1995new} algorithm is used. The algorithm is integrated with the software module of the SP and the search space is defined within a range that meets the hardware constraints. The search space of the optimal hyper-parameters is observed by using 50 particles randomly initialized within the predefined range, and the algorithm runs over 100 iterations. The optimization is applied only for the SP network and the evaluation of the SP for any given hyper-parameters is performed using the SDR classifier, the highest accuracy of which represents the optimal point. For three separated runs and different benchmarks, we run PSO to get the optimal hyper-parameters that result in the highest recognition accuracy. The hyper-parameters that are included in the search space are: the number of winning mini-columns which impacts the SP sparsity level, $minOverlap$ which influences the sparsity level and SP noise robustness, permanence parameters which determine the learning and forgetting rate, and the proximal segment size which controls each mini-column overlap level with the input space.~\tb{para_table} lists the hyper-parameters search space and the optimal values for each benchmark.

\begin{table}[h!tb]
\small
\caption{HTM-SP parameters for the suites benchmarks.}
\label{para_table}
\setlength\tabcolsep{3 pt}
\renewcommand{\arraystretch}{1.3}
\begin{center}
\begin{tabular}{|l|c|c|c|}
  \hline
\rowcolor{Gray}
\textbf{Parameter} & \textbf{Range} & \textbf{MNIST} & \textbf{YaleFaces}  \\ \hline
  Number of winning mini-columns    & 5-40    & 40   & 16   \\ \hline
  MinOverlap                        & 2-25    & 3    & 20    \\ \hline
  Permanence threshold              & 0-0.8   & 0.52 & 0.5  \\ \hline
  Permanence increment ($P^+$)      & 0-0.2   & 0.01 & 0.1 \\ \hline
  Permanence decrement ($P^-$)      & 0-0.2   & -0.01& -0.15 \\ \hline
  Proximal segment size             & 10-500   & 32   & 250   \\ \hline
\end{tabular}
\end{center}
\end{table}

%========================= SP Evaluation Metrics ==================
\section{Spatial Pooler Evaluation Metrics} \label{sp_metrics}
The performance of the SP is evaluated using the metrics defined in~\cite{cui2017htm}, the sparseness and entropy, which reflect the sparsity level of encoding and efficient use of the available resources (mini-columns) in the encoding process. The dataset used during the evaluation is suggested in~\cite{cui2017htm} and consists of a set of random vectors with sparsity level varies between 2\% to 20\% (it is called random dataset in the rest of the paper). In the following subsections, each evaluation metric is discussed in detail.

\subsection{Sparseness}
Sparseness ($\eta$) defines the activity of the mini-columns overtime to ensure the fixed sparsity level in the SDR produced by the SP. Given a SP SDR output vector ($\vec{\Lambda}$), its sparsity level can be measured by dividing the Hamming weight of $\vec{\Lambda}$, $|\vec{\Lambda}|_1$, by the mini-columns count ($n_c$) in the HTM region, and as given in \eq{sparsity_level}
\begin{equation}\label{sparsity_level}
\eta  = \dfrac {|\vec{\Lambda}|_1}{n_c} \times 100
\end{equation}

\begin{figure}[h!tb]
\centering
\subfigure[]{\includegraphics[width=50mm, height=40mm]{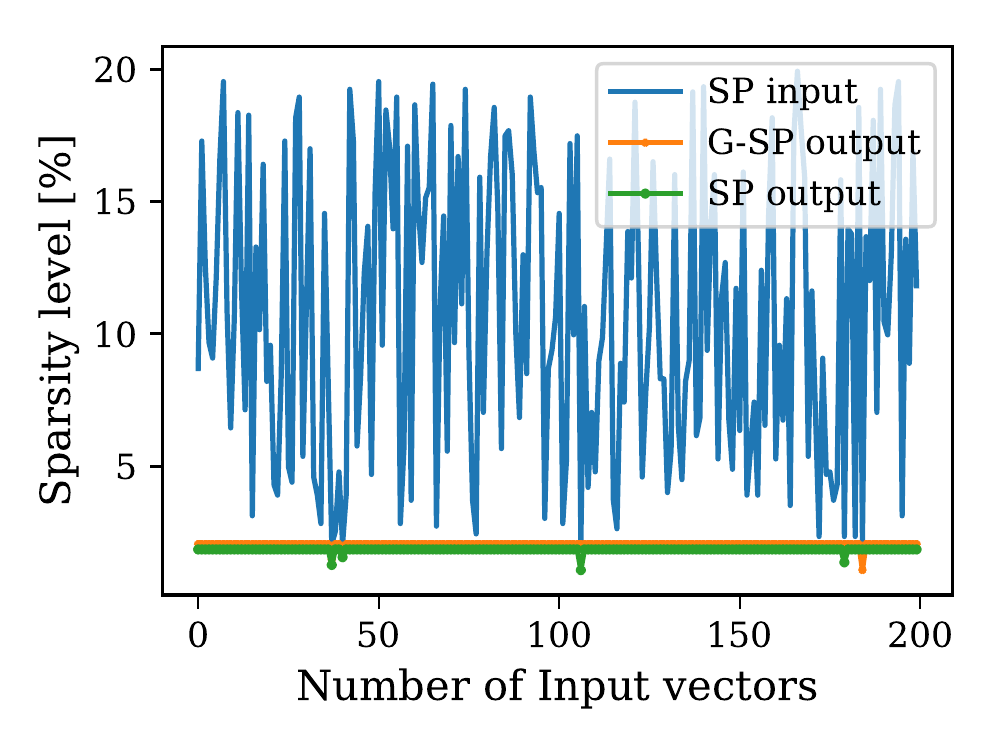}}
\hspace*{1em}
\subfigure[]{\includegraphics[width=50mm, height=40mm]{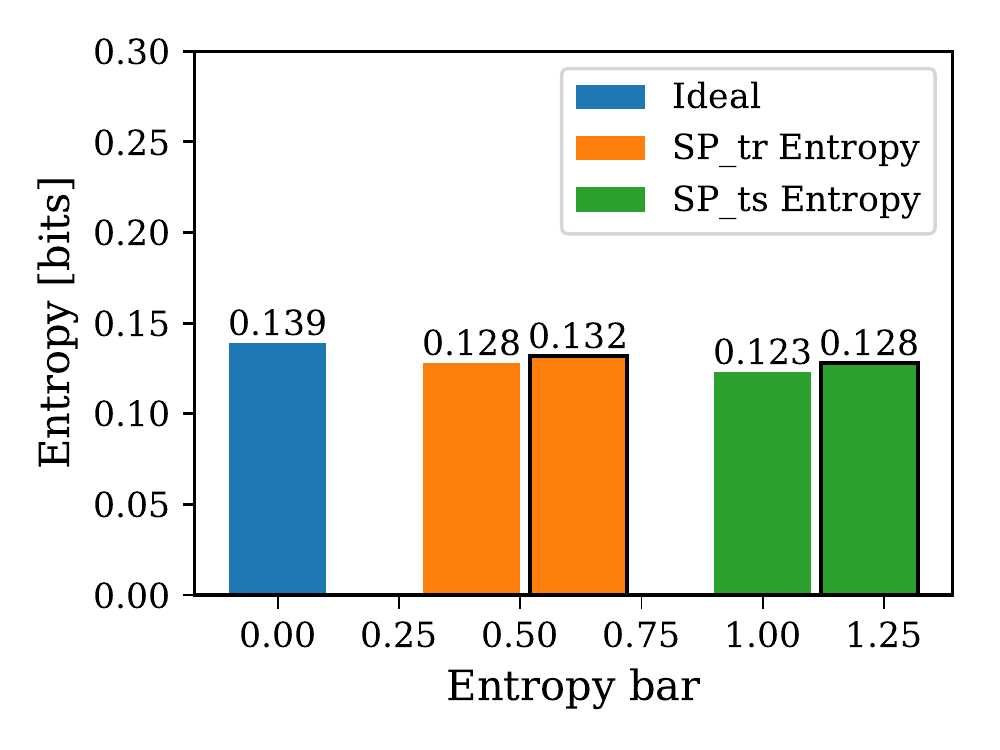}}
\caption{(a) Sparsity level of the input samples and their corresponding SP representations, (b) The average entropy across all the mini-columns for the random dataset when the learning is enabled and disabled.}
\label{EntropyX}
\end{figure}
% check the x-axis of figure(a)

\fig{EntropyX}-(a) illustrates the sparsity level for 200 input samples before and after being treated by the SP. It can be observed that the wide variability in the sparsity level in the input has been significantly reduced to almost~$\approx$2\%. However, due to the fact that during this experiment the $minOverlap$ is set to 4, having 2\% of the sparsity level is not guaranteed as each mini-column should have at least 4 active proximal connections to get involved in the input representation. This issue can be noticed in the negative spikes of the green line which is an indicator of a degradation in the sparsity level of SP output (below 2\%). After adding the neurogenesis mechanism, this issue is actually reduced as seen in orange line (G-SP). This is because the mini-columns that were inactive are replaced with new ones with a higher likelihood to be active.

\subsection{Entropy}
The entropy metric quantifies whether the SP uses all the mini-columns in the region or not. This metric can be computed by summing up the binary entropy function of each mini-column in the HTM region and as given by \eq{entro}~\cite{cui2017htm}, where $E$ is the mean SP entropy and $P(a_j)$ indicates the average activation frequency of the $j^{th}$ mini-column across $n_m$ inputs.~\fig{EntropyX}-(b) illustrates the average entropy\footnote{The mean entropy increases when the mini-columns are equally activated due to the SP fixed sparsity constraints~\cite{cui2017htm}.} of the SP for the random dataset when the learning is enabled and disabled. It can be noticed that enabling the training in the network leads to improve the entropy from 0.123 bits/mini-column to 0.128 bits/mini-column (the maximum possible entropy that can be achieved for the same network setup is 0.139 bits/mini-column). This means that the fraction of mini-columns that are not involved in representing the input before learning became much more active after training. Furthermore, when the neurogenesis mechanism is enabled, further improvement in the entropy has been achieved for both training and testing results (bars with black boxes). This attributes to the continuous mini-columns renewing as a function of their activities.
\begin{equation}
E = \frac{\sum\limits^{n_c}_{j=1}-P(a_j)log_2P(a_j)-(1 - P(a_j))log_2(1-P(a_j))}{n_c}
\label{entro}
\end{equation}
\begin{equation}
P(a_j)  = \frac{1}{n_m} \sum^{n_m}_{t=1} {\Lambda^t_j}
\label{activity}
\end{equation}

%@@@@@@@@@@@@@@@@@@@@@@@@@@@@@@@@@@@@@@@@@@@@@@@@@@@@@@@@@@@@@@@@@@@@@@@
\section{Experimental Results} \label{exper_results}
\subsection{Image Recognition}
SP performance is evaluated on the image recognition task which is conducted using several benchmarks including MNIST\footnote{MNIST is the standard benchmark for hand-written images. It has grayscale images of 28x28 pixels associated with 10 classes for numbers from 0 to 9. The images are split into 60,000 training examples and 10,000 testing examples.}
%fashion MNIST\footnote{Fashion MNIST is the Zalando's article images. Just as in MNIST, the dataset has 10 classes and all the images are in grayscale. The training and testing are 60,000 and 10,1000 examples, respectively.}
and Yale faces\footnote{Yale faces dataset contains 165 grayscale images corresponding to 15 subjects, 11 images each. The images are taken under different conditions and variations including illumination effects, facial expression, etc. In this work, the set is randomly split into training and testing examples, where the training examples contain 8 samples from each subject and 3 samples are used for testing.} datasets. For MNIST, all the images are resized from their original size, 28x28, to 32x32 pixels. Then, all images are binarized by thresholding prior to introduce them to SP. The same process is applied for Yale faces, but here the images are cropped using the face detection Open-CV python library prior to the resizing. The binarization, here, is performed using adaptive thresholding to preserve most image details during the conversion process.

In separate experiments, the data is introduced to the SP as training and testing sets. When the training set is introduced, the SP learns the feed-forward input (images) in an unsupervised fashion. Then, its output is relayed to an SDR classifier implemented by the winner-take-all circuit. The SDR classifier, here, is trained in a supervised fashion using the delta rule. Then, the learning is disabled in both SP and SDR classifier and the testing set is presented to the network. The results, shown in~\tb{accuracyTable}, demonstrate that the network is able to classify the SDR representation generated by the SP with a testing accuracy of 90.33\% for MNIST. In case of Yale faces, due to the limited available training samples, the same training set is presented to the network several times and the resulting accuracy for testing, averaged over 10 runs, is 86.86\%.

%It is important to mention here that high level simulation model that we used to model HTM network has accounted memristor device variability and cycle-to-cycle write variation. The device variability here is confined to device resistance range which is emulated a variation in the weight limit range, while the cycle-to-cycle variation is captured by incorporating undesired noise in the learning process.

In an attempt to compare our results with previous implementations of the SP, for MNIST, it is found that although our network is smaller in size, it still offers a comparable accuracy to other implementations. In case of our previous work, in which 100 mini-columns are used, the high accuracy mainly attributes to the use of high-performance classifier, support vector machine (SVM). When it comes to other sparsity classifiers such as the locally competitive algorithm (LCA), SP+SDR classifier still outperform this classifier with $\approx$ 0.33\%. For Yale faces, SP+SDR classifier outperforms the smooth-marginal fisher analysis (S-MFA) and offers higher average accuracy than the SP implementation in~\cite{james2017htm} which did not consider the entire dataset during the training and testing.

\begin{table}[ht]
\caption{Summary of image recognition accuracy for various datasets using HTM-SP and other algorithms.}
\label{accuracyTable}
\small
\centering
\setlength\tabcolsep{3 pt}
\renewcommand{\arraystretch}{1.3}
\begin{threeparttable}
\begin{tabular}{l|c|c|c|c|}
\cline{2-5}
\rowcolor{Gray}& \textbf{Work}& \textbf{No. of columns}& \textbf{Classifier} & \textbf{Accuracy (\%) $\pm$ STD}\\ \hline
\multicolumn{1}{|c}{\multirow{7}{2mm}{\rotatebox{90}{\parbox{\linewidth}{MNIST}}} }  &
\multicolumn{1}{ |c| }{F-HTM~\cite{streat2016non}} & 784 & SP + SVM (Linear kernel) & 91.98     \\ \cline{2-5}
\multicolumn{1}{ |c  }{}
& \multicolumn{1}{ |c| }{Memristive-LCA \cite{woods2015synaptic}} & 300 & LCA & 90.0     \\ \cline{2-5}
\multicolumn{1}{ |c  }{}  &
\multicolumn{1}{ |c| }{Digital-HTM~\cite{htm_digital}} & 100 & SP + SVM (RBF kernel) & 91.16     \\ \cline{2-5}
\multicolumn{1}{ |c  }{}  &
\multicolumn{1}{ |c| }{Crossbar HTM~\cite{truong2018spatial}\tnotex{tnote:r3}} & 1024 & SP+X & $\approx$90.5     \\ \cline{2-5}
\multicolumn{1}{ |c  }{}  &
\multicolumn{1}{ |c| }{This work} & 484 & SP+SDR classifier & 90.33 $\pm$ 0.17\tnotex{tnote:r4}      \\ \cline{1-5}
\multicolumn{1}{|c}{\multirow{5}{2mm}{\rotatebox{90}
{\parbox{\linewidth}{YaleFaces}}}} &
\multicolumn{1}{ |c| }{Memristor HTM~\cite{james2017htm}\tnotex{tnote:r2}} & - & SP+XOR classifier & 86.67   \\ \cline{2-5}
\multicolumn{1}{ |c  }{}                        &
\multicolumn{1}{ |c| }{Smooth-MFA\cite{cai2007learning}\tnotex{tnote:r1}} & - & S-MFA & 81.1  \\ \cline{2-5}
\multicolumn{1}{ |c  }{}                        &
\multicolumn{1}{ |c| }{This work} & 1024 & SP+SDR classifier & 86.86 $\pm$ 3.82\tnotex{tnote:r4}   \\ \cline{1-5}
\end{tabular}
\begin{tablenotes}
\item\label{tnote:r1} Software implementation.
\item\label{tnote:r2} In \cite{james2017htm}, the SP parameters to achieve the aforementioned accuracy are not included. Also, the reported average accuracy is when the network is tested separately only on emotions, light conditions, facial expressions portions of the dataset.
\item\label{tnote:r3} In~\cite{truong2018spatial}, 95\% classification accuracy is reported for using 4096 mini-columns, but it is not mentioned if this is for a hardware implementation. Furthermore, the authors did not mention the type of classifier used with the SP, for this reason, we do denote it by X.
\item\label{tnote:r4} The high-level simulation model that we developed to model HTM network has accounted for memristor device variability and cycle-to-cycle write variation. The device variability here is confined to device resistance range which is emulated as a variation in the weight range, while the cycle-to-cycle write variation is modeled by adding noise to the learning rule.
\end{tablenotes}
\end{threeparttable}
\end{table}

\subsection{Noise Robustness}
\begin{figure}[h!tb]
\centering
\subfigure[]{\includegraphics[width=50mm, height=40mm]{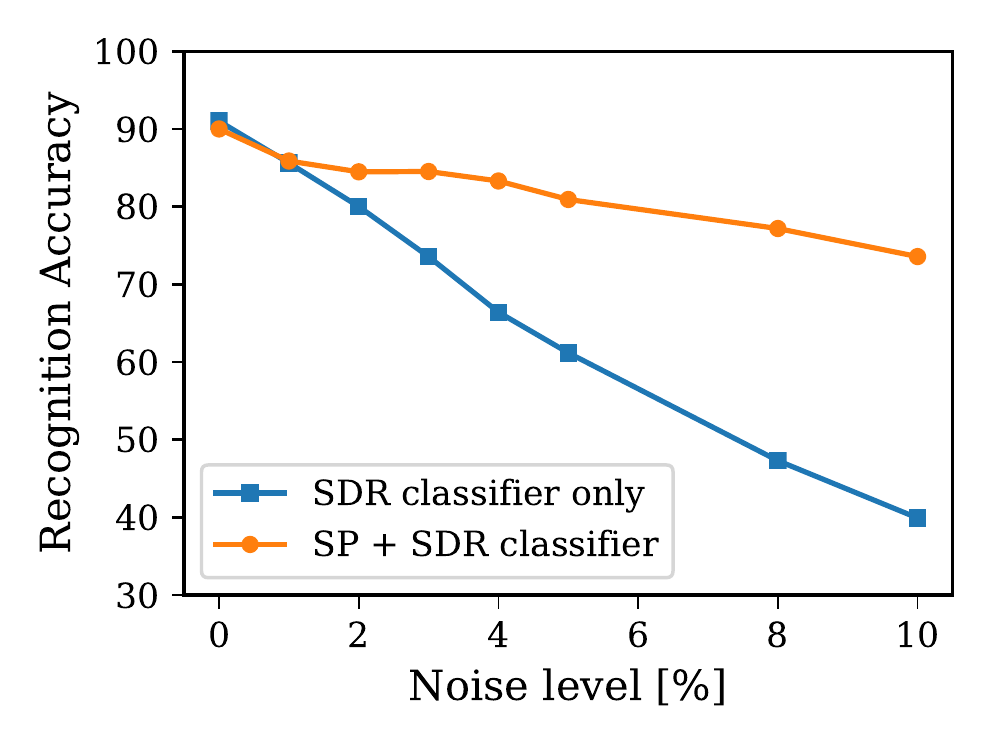}}
\hspace*{1em}
\subfigure[]{\includegraphics[width=50mm, height=40mm]{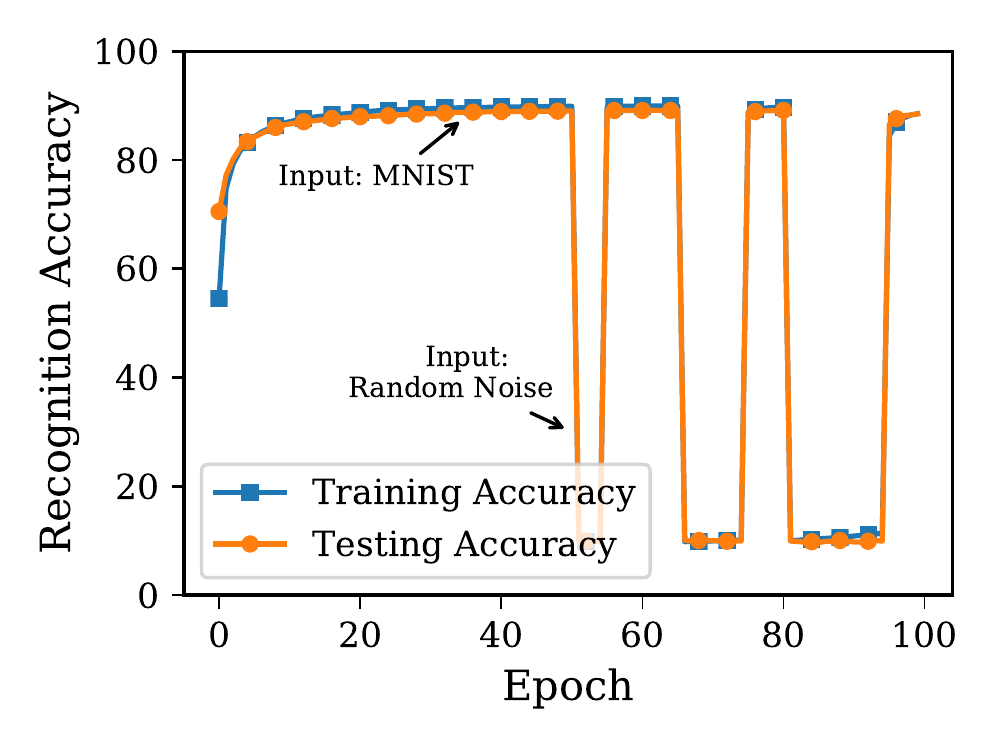}}
\caption{(a) Recognition accuracy of MNIST dataset classified with SDR classifier and SP+SDR classifier in the presence of a noise level range between 0\% and 10\%. (b) Noise robustness of the SP+SDR classifier when presenting MNIST dataset as a stream of data mediated by noisy information.}
\label{accur_noise}
\end{figure}

In order to quantify the noise robustness of the SP+SDR classifier for image recognition applications, two experiments are performed. The first of which involves classifying MNIST images in the presence of noise and the second one involves interrupting the training process by injecting random SDRs. For the first experiment, the SP+SDR classifier is trained on clean training MNIST images and tested with noisy test images. The noise here is added by flipping the image pixels randomly. The noise level is defined by the percentage of the flipped pixels in an image. For a noise level ranging between 0\% to 10\%, both the SDR classifier ($\equiv$ softmax classifier) and SP+SDR classifier are tested separately.~\fig{accur_noise} demonstrates the drop in recognition accuracy as both classifiers are tested with corrupted MNIST images with various noise levels. It can be observed that the SP+SDR classifier was able to handle the noise with a graceful degradation in accuracy in comparison to the SDR classifier which its accuracy dropped to 37.7\%, when 10\% noise is added to the images.

In the second experiment, the SP is used to generate the SDR representations of MNIST training and testing sets. Then, the SDRs representations are presented to the SDR classifier during the training and testing processes, where one cycle of training and testing is considered as one epoch. MNIST SDRs are presented to the SDR classifier for 100 epochs mediated by noise injection. The noise here consists of a set of random SDR vectors with sparsity level similar to that in MNIST SDR vectors as generated by the SP. 10,000 noise vectors are generated and injected in parts after the SDR classifier settles to a reasonable accuracy level ($\approx$ 90.38). Between epochs 50-55, random 1000 SDR vectors are presented to the SDR classifier, between epochs 65-75, 5000 random SDR vectors are used, and between epochs 80-95, 10,000 random vectors are used. \fig{accur_noise}-(b) illustrates the drop in recognition accuracy when the data streams are replaced by a stream of noisy vectors and the fast recovery after the noisy vectors are removed. The fast recovery, here, attributes to training the SDR classifier on sparse inputs which makes the likelihood of adjusting the critical connections (i.e. weights) less likely. This, consequently, shows that the degradation in classifier performance, even after removing the noise vectors, is almost negligible.

%%%%%%%%%%%%%%%
\subsection{Power Consumption and Area}
\begin{figure}[h!tb]
\centering
\subfigure[]{\includegraphics[width=50mm, height=40mm]{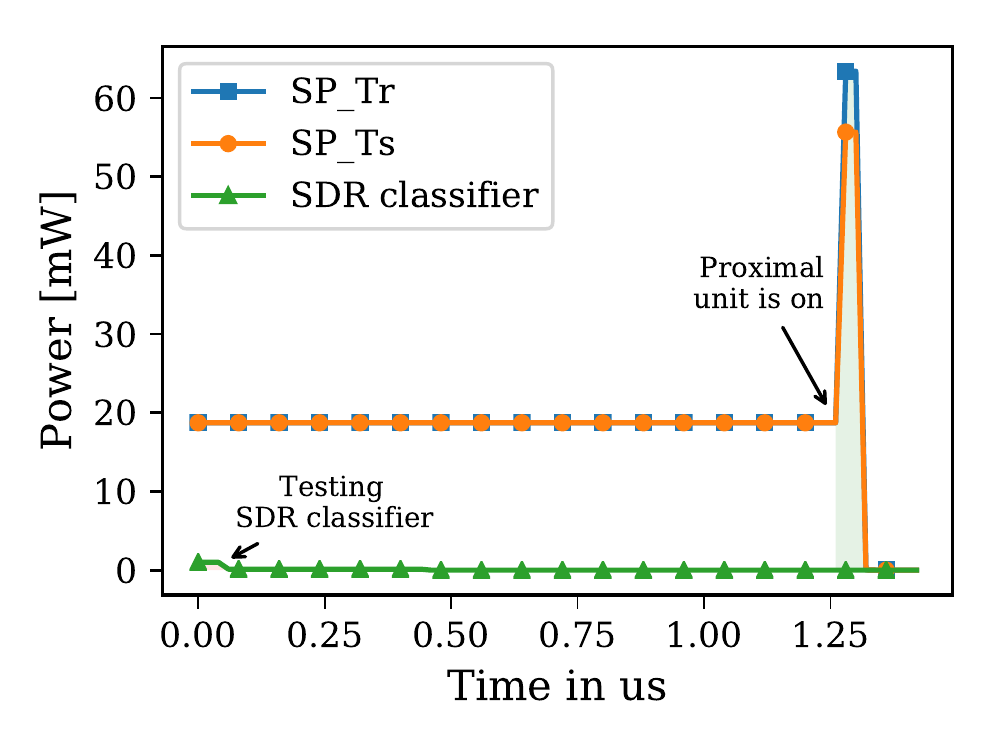}}
\hspace*{1em}
\subfigure[]{\includegraphics[width=50mm, height=40mm]{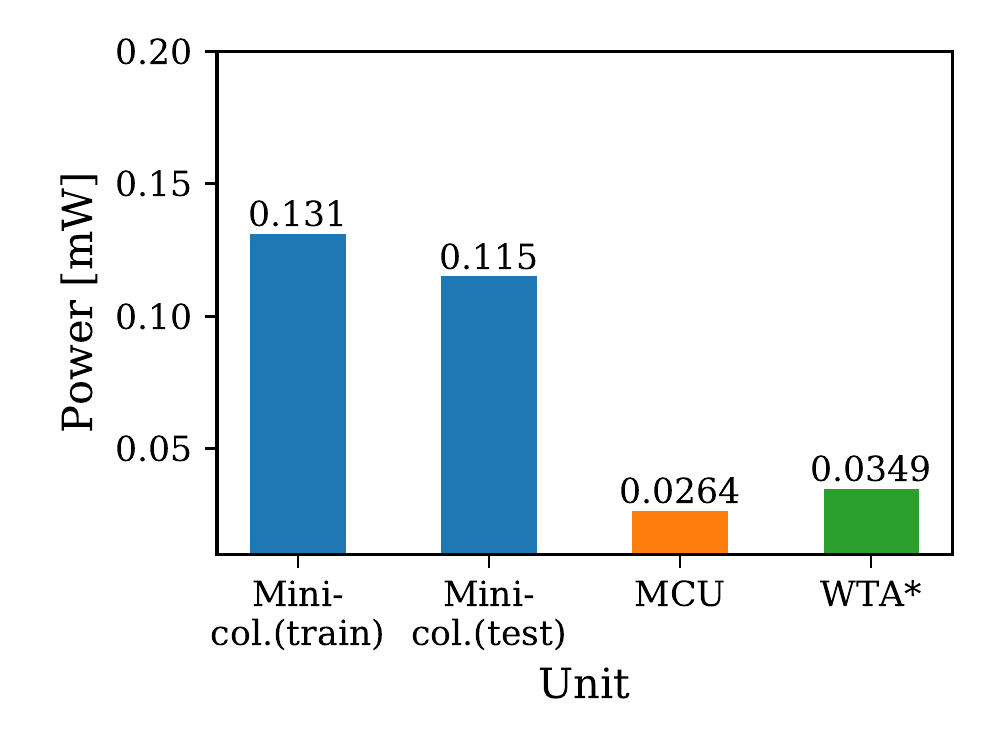}}
\caption{(a) The average total power consumption of the SDR classifier and SP during training and testing when the network is used to classify MNIST dataset ($n_c$=484, $n_s$=32). (b) Average total power consumption of each unit in the SP (WTA size = 1000 cells).}
\label{power}
\end{figure}

\fig{power}-(a) shows the total power consumption\footnote{The power consumption of the proposed design is evaluated in Cadence (analog blocks) and Synopsys (digital blocks) for 65nm technology node. The digital units are integrated with a testbench to fetch the training images to the network and their loads are emulated by D-FFs. Initially, we run the circuit for 1ms to capture the circuit switching activity. Then, we use Synopsys tools to measure the power consumption of the digital units. In case of the analog blocks, a crossbar network and mini-column training circuitry with memristor proximal connections serve as a testbed. During the testing, we consider an input voltage of 0.9v, whereas 1.2v is used during the training. It is important to mention here that while measuring the power consumption, all the memristors are replaced by resistors and we consider the worst case scenario, (all crossbar memristors are set to low resistance state). Then, the power consumption is estimated using Cadence ADE tools by covering all the input combinations and averaging the results. For the area estimation, we used Synopsys tools to estimate the area of the digital units and the analog units area is estimated from the physical layout designed in Cadence.
%To get more precise estimation, the digital units I/O are loaded with a D-FF  All the digital units power consumption is evaluated in Synpo
} of the SP and the SDR classifier during the training and testing processes when the network is used to recognize MNIST digits, running at 50MHz. For a single iteration\footnote{Iteration in this context refers to the time starting from fetching an image to the network until it gets recognized. It also includes the training time of the SP and the classifier.}, initially, the power consumption resides at 18.72mW until each mini-column activates its corresponding proximal unit. During the training, the proximal connections are driven by the training voltage (1.2v) rather than the testing voltage (0.9v) to adjust the strength of the connections leading to greater power consumption. However, since the use of a proximal unit takes 2 clock cycles, we do not see this abrupt increase in the power last long. It is important to mention here that the SP and SDR classifier are working simultaneously and in a pipelined fashion. Thus, when the SDR classifier is used for inference, there is around 0.994mW of power consumption which then degrades to 0.121mW during the training as only one crossbar column is trained in 2 clock cycles. Finally, when the SP communicates with the testbench according to the hand-shake protocol to receive a new patch, the network experiences a sudden drop in power (last few cycles) as most system units are disabled. In~\fig{power}-(b), the total average power consumption of each unit in the SP design is illustrated. From here, we can see that the power consumption of the mini-columns is dominating the network. It can be reduced further by lowering the operating frequency and if memristors with higher resistance are used. Finally, the network (SP+SDR classifier) setup used to classify the MNIST dataset consumes an average total power of 18.47mW with an area estimate of 0.513mm$^2$.

%initially, the power consumption resides at 18.72mW until the training of proximal connections occurs. During the training, the proximal connections are driven by the training voltage (1.2v) rather than (0.9v) to adjust the strength of the connections. However, since training a mini-column takes 2 clock cycles, we do not see this abrupt increase in the power last long. It is important to mention here that the SP and SDR classifier are working simultaneously and in a pipelined fashion. Thus, when the SDR classifier is used for inference, there is around 0.994mW of power consumption which then degrades to 0.121mW during the training as only one crossbar column is trained in two clock cycles. Finally, when the SP communicates with the testbench according to the hand-shake protocol to receive a new patch, the network experiences a sudden drop in power (last few cycles) as most system units are disabled. In~\fig{power}-(b), the total average power consumption of each unit in the SP design is illustrated. From here, we can see that the power consumption of the mini-columns dominating the network. It can be reduced further at lower operating frequency and if memristors with higher resistance are used. Finally, the network (SP+SDR classifier) setup used to classify MNIST dataset consumes an average total power of 18.47mW with an area estimate of 0.513mm$^2$.

\section{Conclusions} \label{conclusion}
This paper proposes a memristor-based architecture for the HTM spatial pooler and its SDR classifier for mobile devices and energy constrained platforms. The proposed design enables high-speed computations, low power consumption, and reconfigurability, all in a single entity that has the capability to recognize images even in the presence of noise. The proposed design is implemented using 65nm technology node and verified using various datasets including MNIST (accuracy 90.33\%) and Yale faces (accuracy = 86.86\%). It is found that the network exhibits a strong robustness to noise especially from the classifier side as it is trained with SDR representations. Furthermore, during the power consumption analysis, it is observed that the power consumption while using the proximal unit is approximately tripled. However, limiting the use of this unit to 2 clock cycles significantly reduces the overall network power consumption to 18.47mW.

%This paper proposes a memristor-based architecture for the HTM spatial pooler and its SDR classifier for mobile devices and energy constraint platforms. The proposed design enables high-speed computations, low power consumption, and reconfigurability, all in a single entity that has the capability to recognize images even in the presence of noise. The proposed design is implemented using 65nm technology node and verified for various datasets including MNIST (accuracy 90.45\%) and Yale faces (accuracy = 84.11\%). When the network is tested in the presence of noise, it is found that the network exhibits a strong robustness for noise especially from the classifier side as it is trained with SDR representations. Furthermore, during the power consumption analysis, it is observed that the power consumption while training the proximal connections doubled almost three times. However, limiting the training process to 2 clock cycles significantly reduced the overall network power consumption to be 18.72mW.

%The future work may involve exploring the hardware implementation of the TM and testing the network on time-series data.

%This paper investigates the memristor-based mixed-signal implementation of the HTM-SP and the SDR classifier as a single entity. It also explores the effect of including structure plasticity mechanism such as the neurogenesis on network performance.
%@@@@@@@@@@@@@@@@@@@@@@@@@@@@@@@@@@@@@@@@@@@@@@@@@@

\begin{acks}
The authors would like to thank the members of the Neuromorphic AI research Lab at RIT for their support and critical feedback. The authors also would like to thank the reviewers for their time and extensive feedback to enhance the quality of the paper.
\end{acks}

% Bibliography

\end{document}